\documentclass[twocolumn,tighten]{aastex61}
\usepackage{rotating}

\hypersetup{linkcolor=blue,citecolor=blue,filecolor=cyan,urlcolor=blue}

\newcommand{\wcen}{$\omega$~Cen}
\newcommand{\kms}{${\rm km}\,{\rm s}^{-1}$}
\newcommand{\masyr}{${\rm mas}\,{\rm yr}^{-1}$}
\newcommand{\maspx}{${\rm mas}\,{\rm pixel}^{-1}$}
\newcommand{\dmacd}{$\Delta \mu_\alpha \cos \delta$}
\newcommand{\dmd}{$\Delta \mu_\delta$}
\newcommand{\mtan}{$\mu_{\rm tan}$}
\newcommand{\mrad}{$\mu_{\rm rad}$}
\newcommand{\stan}{$\sigma_{\rm tan}$}
\newcommand{\sm}{$\sigma_\mu$}
\newcommand{\srad}{$\sigma_{\rm rad}$}
\newcommand{\mv}{$m_{\rm F606W}$}

\newcommand{\mvi}{$m_{\rm F606W}-m_{\rm F814W}$}
\newcommand{\mjh}{$m_{\rm F110W}-m_{\rm F160W}$}

\received{December 4, 2017}
\revised{December 14, 2017}
\accepted{December 15, 2017}

\shorttitle{Internal kinematics}
\shortauthors{Bellini et al.}

\begin{document}

\title{The \textit{HST} large programme on $\omega$~Centauri
  -- II. Internal kinematics}

\correspondingauthor{Andrea Bellini}
\email{bellini@stsci.edu}

\author[0000-0003-3858-637X]{Andrea Bellini} \affil{Space Telescope
  Science Institute, 3700 San Martin Drive, Baltimore, MD 21218, USA}

\author[0000-0001-9673-7397]{Mattia Libralato} \affil{Space Telescope
  Science Institute, 3700 San Martin Drive, Baltimore, MD 21218, USA}

\author[0000-0003-4080-6466]{Luigi R. Bedin} \affil{Istituto Nazionale
  di Astrofisica, Osservatorio Astronomico di Padova, Vicolo
  dell'Osservatorio 5, Padova I-35122, Italy}

\author[0000-0001-7506-930X]{Antonino P. Milone} \affil{Dipartimento
  di Fisica e Astronomia ``Galileo Galilei'', Universit\`a di Padova,
  Vicolo dell'Osservatorio 3, Padova I-35122, Italy}

\author[0000-0001-7827-7825]{Roeland P. van der Marel} \affil{Space
  Telescope Science Institute, 3700 San Martin Drive, Baltimore, MD
  21218, USA} \affil{Center for Astrophysical Sciences, Department of
  Physics \& Astronomy, Johns Hopkins University, Baltimore, MD 21218, USA}

\author[0000-0003-2861-3995]{Jay Anderson} \affil{Space Telescope
  Science Institute, 3700 San Martin Drive, Baltimore, MD 21218, USA}

\author[0000-0003-3714-5855]{D\'aniel Apai} \affil{Department of
  Astronomy and Steward Observatory, The University of Arizona, 933
  North Cherry Avenue, Tucson, AZ 85721, USA} \affil{Lunar and
  Planetary Laboratory, The University of Arizona, 933 North Cherry
  Avenue, Tucson, AZ 85721, USA}

\author[0000-0002-6523-9536]{Adam J. Burgasser} \affil{Center for
  Astrophysics and Space Science, University of California, San Diego,
  La Jolla, CA 92093, USA}

\author{Anna F. Marino} \affil{Research School of Astronomy \&
  Astrophysics, Australian National University, Canberra, ACT 2611,
  Australia}

\author{Jon M. Rees} \affil{Department of Astronomy and Steward
  Observatory, The University of Arizona, 933 North Cherry Avenue,
  Tucson, AZ 85721, USA}

\begin{abstract}

  In this second installment of the series, we look at the internal
  kinematics of the multiple stellar populations of the globular
  cluster $\omega$~Centauri in one of the parallel \textit{Hubble
    Space Telescope} (\textit{HST}) fields, located at about 3.5
  half-light radii from the center of the cluster. Thanks to the over
  15-year-long baseline and the exquisite astrometric precision of the
  \textit{HST} cameras, well-measured stars in our proper-motion
  catalog have errors as low as $\sim 10\,\mu$as\,yr$^{-1}$, and the
  catalog itself extends to near the hydrogen-burning limit of the
  cluster. We show that second-generation (2G) stars are significantly
  more radially anisotropic than first-generation (1G) stars. The
  latter are instead consistent with an isotropic velocity
  distribution.  In addition, 1G have excess systemic rotation in the
  plane of the sky with respect to 2G stars.  We show that the six
  populations below the main-sequence (MS) knee identified in our
  first paper are associated to the five main population groups
  recently isolated on the upper MS in the core of
  cluster. Furthermore, we find both 1G and 2G stars in the field to
  be far from being in energy equipartition, with $\eta_{\rm
    1G}=-0.007\pm0.026$ for the former, and $\eta_{\rm
    2G}=0.074\pm0.029$ for the latter, where $\eta$ is defined so that
  the velocity dispersion $\sigma_\mu$ scales with stellar mass as
  $\sigma_\mu \propto m^{-\eta}$.  The kinematical differences
  reported here can help constrain the formation mechanisms for the
  multiple stellar populations in $\omega$~Centauri and other globular
  clusters.  We make our astro-photometric catalog publicly available.

\end{abstract}


\section{Introduction} \label{sec:1}

\begin{table*}[t!]
\caption{List of \textit{HST} observations of field F1.\label{tab:1}}
\centering
{
\begin{tabular}{lllll}
\hline\hline
Filter& Exposures& Program ID& PI& Epoch\\
\hline
\multicolumn{5}{c}{ACS/WFC (Epoch 1)}\\
\hline
F606W& $2\times 1300\,{\rm s} +2\times 1375\,{\rm s}$&9444& King, I. R.&2002/07/03\\
F814W& $2\times 1340\,{\rm s} +2\times 1375\,{\rm s}$&9444& King, I. R.&2002/07/03\\
\hline
\multicolumn{5}{c}{ACS/WFC (Epoch 2)}\\
\hline
F606W& $2\times 1285\,{\rm s} +2\times 1331\,{\rm s}$&10101&King, I. R.&2005/12/24\\
F814W& $4\times 1331\,{\rm s}$&10101&King, I. R.&2005/12/24\\
\hline
\multicolumn{5}{c}{WFC3/UVIS (Epoch 3)}\\
\hline
F275W&$4\times 1328\,{\rm s}$&14118& Bedin, L. R.&2015/08/23--26\\
F336W&$4\times 1230\,{\rm s}$&14118& Bedin, L. R.&2015/08/23--26\\
F438W&$4\times 98\,{\rm s}$&14118& Bedin, L. R.&2015/08/22--26\\
F606W&$2\times 99\,{\rm s} + 2\times 1255\,{\rm s} +2\times 1347\,{\rm s}$&14118& Bedin, L. R.&2015/08/22--23\\
F814W&$2\times 98\,{\rm s} + 2\times 1253\,{\rm s} +2\times 1345\,{\rm s}$&14118& Bedin, L. R.&2015/08/20--21\\
\hline
\multicolumn{5}{c}{WFC3/IR (Epoch 3)}\\
\hline
F110W&$7\times 142\,{\rm s} + 14\times 1302\,{\rm s}$&14118&Bedin, L. R.&2015/08/19--24\\
F160W&$7\times 142\,{\rm s} + 14\times 1302\,{\rm s}$&14118&Bedin, L. R.&2015/08/24--26\\

\hline
\multicolumn{5}{c}{WFC3/UVIS (Epoch 4)}\\
\hline
F275W&$4\times 1229\,{\rm s}$&14662& Bedin, L. R.&2017/08/19--20\\
F336W&$4\times 1143\,{\rm s}$&14662& Bedin, L. R.&2017/08/19--20\\
F438W&$3\times 95\,{\rm s}+ 1\times 104\,{\rm s}$&14662& Bedin, L. R.&2017/08/19--20\\
F606W&$2\times 104\,{\rm s} + 2\times 1172\,{\rm s} +2\times 1252\,{\rm s}$&14662& Bedin, L. R.&2017/08/19\\
F814W&$2\times 104\,{\rm s} + 2\times 1172\,{\rm s} +2\times 1252\,{\rm s}$&14662& Bedin, L. R.&2017/08/19\\
\hline
\multicolumn{5}{c}{WFC3/IR (Epoch 4)}
\\
\hline
F110W&$7\times 142\,{\rm s} + 14\times 1202\,{\rm s}$&14662&Bedin, L. R.&2017/08/20--24\\
F160W&$7\times 142\,{\rm s} + 14\times 1202\,{\rm s}$&14662&Bedin, L. R.&2017/08/24--26\\
\hline
\end{tabular}}
\end{table*}

Galactic globular clusters (GCs) have long been assumed to be the best
examples of simple stellar populations (SSPs), i.e., made of stars of
different masses but at the same age and chemical composition.  This
paradigm was first challenged in the late sixties by the massive GC
$\omega$\,Centauri (NGC\,5139, hereafter \wcen).  Photometrically,
\citet{1966ROAn....2....1W} found a sizable broadening of the
red-giant branch (RGB), later solidly confirmed by
\citet{1999Natur.402...55L} and \citet{pancino00}.  Significant
chemical anomalies among RGB stars were initially detected
spectroscopically by \citet{1967RGOB..128..255D}, and later confirmed
by many authors (e.g., \citealt{2011ApJ...731...64M} and references
therein).

However, the divide between the traditional picture of SSP GCs and the
well established presence of multiple stellar populations (mPOPs)
within GCs was the discovery (\citealt{anderson97}) and confirmation
(\citealt{2004ApJ...605L.125B}) that the \textit{unevolved} stars
along the main sequence (MS) of \wcen\ split in at least two distinct
groups.  The recent discoveries of mPOPs in \textit{formally all}
Milky-Way GCs has dramatically increased research into formation,
evolution and populations in these systems:\ \wcen\ is no longer the
exception among GCs, but rather just the most extreme case
(\citealt{2017MNRAS.464.3636M} and references therein).

With the continuous development of reduction techniques, detection
methods and observing strategies, each of the mPOP groups identified
by \citet{2004ApJ...605L.125B} in \wcen\ has been further divided into
sub-groups, and two new main population groups were discovered,
bringing the current total number of mPOPs in \wcen\ to at least 15
(\citealt{2017ApJ...842....6B, 2017ApJ...842....7B,
  2017ApJ...844..164B, 2017MNRAS.464.3636M}).

\begin{figure*}[p!]
\centering
\includegraphics[width=1.01\textwidth]{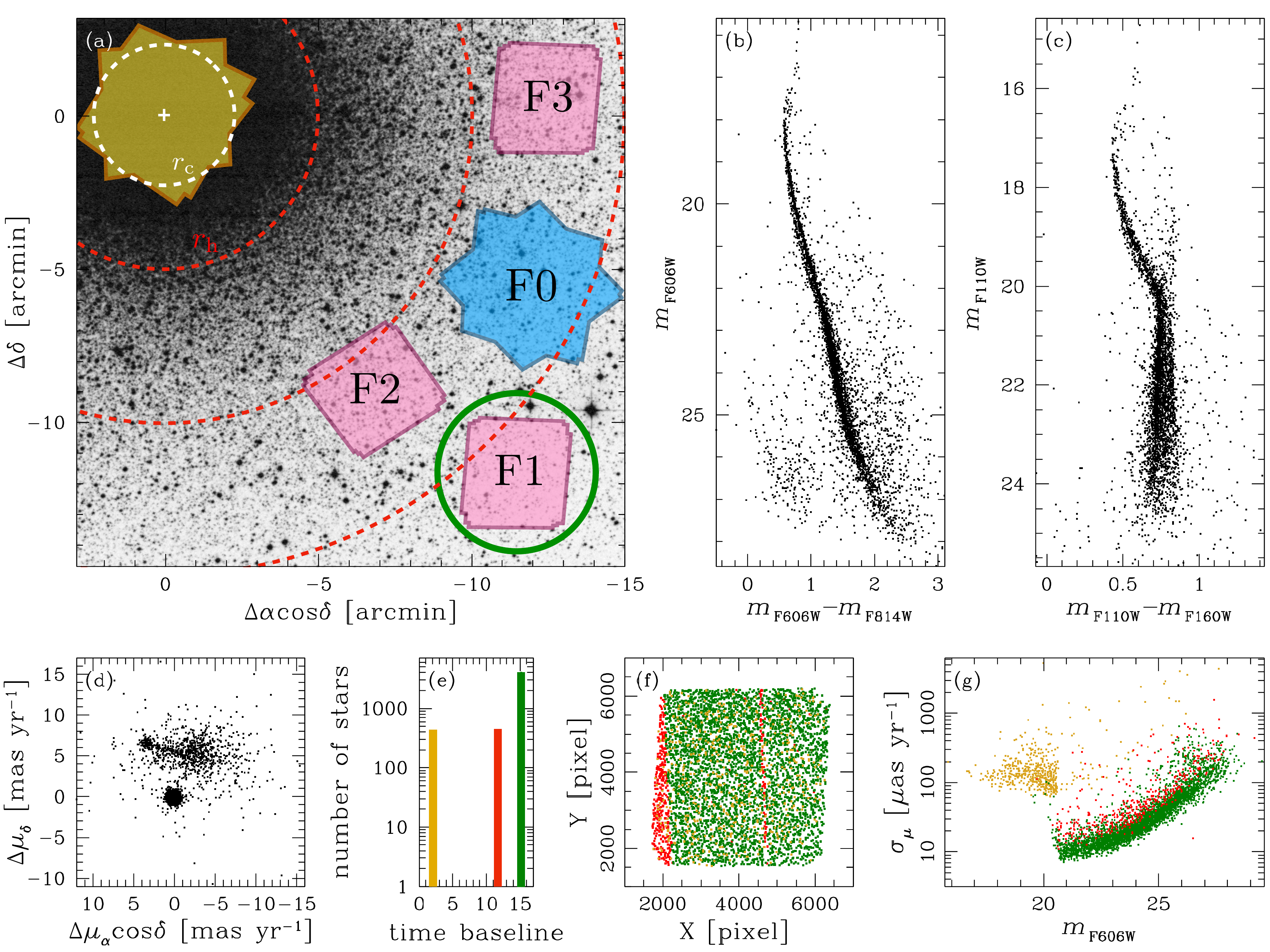}
\caption{(a) Outlines of the fields of GO-14118 + GO-14662 data,
  superimposed on a DSS image of \wcen. Units are in arcmin with
  respect to the cluster's center. The primary ACS/WFC field (F0) is
  in azure, while the three parallel WFC3 fields are in pink. The data
  discussed in this paper come from field F1, which is highlighted by
  a green circle. We also show, in yellow, the central field of
  \citet{2017ApJ...842....6B}. The white and red dashed circles mark
  the core radius ($r_{\rm c}$), the half-light radius ($r_{\rm h}$),
  $2\times r_{\rm h}$ and $3\times r_{\rm h}$ from the center outward,
  respectively. (b) The $m_{\rm F606W}$ versus $m_{\rm F606W}-m_{\rm
    F814W}$ CMD of sources with measured PMs. (c) The $m_{\rm F110W}$
  versus $m_{\rm F110W}-m_{\rm F160W}$ CMD for the same sources.  (d)
  The PM diagram, in units of \masyr. Proper motions are measured
  relative to the bulk motion of the cluster. Cluster members are
  gathered together around (0,0). The lesser clump of points at around
  (+3.5,+6.5) is mainly background galaxies, and will be the subject
  of the next paper in this series (\citealt{libra.p3}). (e) Histogram
  of the temporal baseline used to compute the PM of each
  source. Bright stars are unsaturated only in the short exposures of
  the third and fourth epochs, so that their PM is based on a temporal
  baseline of just two years (yellow bar). The
  longer temporal baseline ($\sim 15.1$\,yr, green bar) is based on
  observations taken in at least the first and the fourth
  epochs. Sources outside the first-epoch FoV but measured in the
  remaining epochs have a temporal baseline of $\sim 11.7$\,yr (red
  bar). (f) The master frame, in units of WFC3/UVIS pixels (40 mas
  pixel$^{-1}$). Sources are color-coded according to the temporal
  baseline used to measure their PM. (g) Proper-motion errors as a
  function of the $m_{\rm F606W}$ magnitude, color-coded as in panels
  (e) and (f). Well-measured stars in the long exposures have typical
  PM errors of about 10 $\mu$as\,yr$^{-1}$.}
\label{fig:1}
\end{figure*}

Understanding how these mPOPs formed and have evolved is now a main
thrust of GC studies (e.g., the dynamical studies of
\citealt{2013ApJ...771L..15R} and \citealt{2015ApJ...810L..13B}).  As
GCs are the oldest objects in the Universe for which reliable ages can
be determined, understanding their formation and evolution is
paramount to understanding the formation and evolution of the Milky
Way itself, and galaxies in general (e.g.,
\citealt{2015ApJ...805..178B}). Among these investigations, the
``\textit{Hubble Space Telescope} (\textit{HST}) large programme of
$\omega$~Centauri'' (GO-14118\,+\,GO-14662, PI: Bedin, L.~R.)  aims at
analyzing the mPOP phenomenon among the faintest white dwarfs (WDs) in
the two cooling sequences of
\wcen\ (\citealt{2013ApJ...769L..32B}). The program is currently
observing a primary field (field F0, see panel (a) of
Fig.~\ref{fig:1}) with the Wide-Field Channel (WFC) of the Advanced
Camera for Surveys (ACS), located about $13\farcm5$ from the cluster's
center, and three surrounding parallel fields (fields F1, F2 and F3)
with both the InfraRed (IR) and the Ultraviolet-VISible (UVIS)
channels of the Wide-Field Camera 3 (WFC3). All the planned data for
the parallel field F1, which was previously observed in 2002 (GO-9444,
PI:\ King, I.~R., see \citealt{2004ApJ...605L.125B}) and in 2005
(GO-10101, PI:\ King, I.~R., see \citealt{2012AJ....144....5K}), have
been observed.

Thanks to the large mass of the cluster ($\sim 4 \times 10^6\,
\mathrm{M}_\odot$, \citealt{2013MNRAS.429.1887D}), we do not expect
mPOPs in the fields at $\gtrsim 17^\prime$, ($\gtrsim 3$ half-light
radii, \citealt{h96}) to be relaxed (e.g., \citealt{dercole2008,
  decressin}), so the fossil information of their initial kinematic
properties should still be observable.  Differences in kinematics
among the mPOPs, coupled with differences in the radial distribution
(e.g., \citealt{2007ApJ...654..915S, 2009A&A...507.1393B}) can be used
to identify precious clues on the formation and evolution of mPOPs in
particular, and of GCs in general.

The long temporal baseline (over 15 years) and the depth available in
all epochs in field F1 enable a detailed study of the internal
kinematics of the mPOPs in \wcen\ through high-precision proper
motions (PMs), which is the main subject of this work.

\section{Data set and reduction} \label{sec:2}

Field F1 has been observed a total of four times by \textit{HST}.
Long ACS/WFC observations in F606W and F814W were taken during the
first two visits, in July 2002 (GO-9444) and in December 2005
(GO-10101). More recently, as part of our \textit{HST} large program,
the field was re-observed in August 2015 (GO-14118) and August 2017
(GO-14662), using both channels of the WFC3. In each of these two
recent epochs, UVIS filters F275W, F336W, F438W (the so-called ``magic
trio'', e.g., \citealt{2015AJ....149...91P}), F606W and F814W, and IR
filters F110W and F160W were utilized. The complete list of
\textit{HST} observations of field F1 is reported in
Table~\ref{tab:1}. A link to the data is provided
here:\ \dataset[10.17909/T9FD49]{http://dx.doi.org/10.17909/T9FD49}.

Panel (a) of Fig.~\ref{fig:1} shows the location of the footprints of
the GO-14118 and GO-14662 fields (F0 to F3) with respect to the center
of the cluster (white cross), superimposed on a DSS
image.\footnote{\url{https://archive.stsci.edu/dss/}} Coordinates in
arcmin with respect to the cluster's center as measured by
\citet{2010ApJ...710.1032A}: (R.A.,Dec.) = (13$^{\rm h}$26$^{\rm
  m}$47.$\!\!^{\rm s}$24,$-$47$^{\circ}$28$^\prime$46$\farcs$45).  The
primary ACS/WFC field (F0) is in azure, while the three parallel WFC3
fields are in pink. As a reference, we also show the central field (in
yellow) analyzed in \citet{2017ApJ...842....6B, 2017ApJ...842....7B,
  2017ApJ...844..164B}. The cluster's core ($r_{\rm c}=1\farcm31$) and
half-light ($r_{\rm h}=2\farcm37$) radii are marked with white and red
dashed circles, respectively. The two outer red circles have a radius
of $2\times r_{\rm h}$ and $3\times r_{\rm h}$, respectively. GO-14118
and GO-14662 fields cover a radial extent from $\sim 2\times r_{\rm
  h}$ to $\sim 4\times r_{\rm h}$. This paper focuses on field F1
(encircled in green), which is the only field for which all exposures
have already been acquired.

The astro-photometric catalogs obtained in \citet[hereafter,
  Paper~I]{2017MNRAS.469..800M} for field F1 were constructed from
the onset with the goal of detecting fine substructures on the
color-magnitude diagram (CMD). Here our goal is to measure fine
substructures on the PM diagram.  Photometry and astrometry make very
different demands on PSF analysis, with photometry more focused on
sums of pixels, whereas for astrometry differences between nearby
pixel values are key. A good PSF model should measure both fluxes and
positions well, and our state-of-the-art reduction techniques allow
just that. However, measurements that might end up being discarded
because of high-precision photometric needs might still be useful for
high-precision astrometric investigations, and vice versa. Selection
procedures play a crucial role in obtaining appropriate stellar
samples for astrometric or photometric studies, and there is generally
no one-size-fits-all solution. For these reasons, we reduced all of
the available exposures of field F1 from scratch, with the goal of
high-precision astrometry right from the start. We will still make use
of the photometry of Paper~I (Sects~\ref{sec:3} and \ref{sec:4}) to
isolate the mPOPs of the cluster.

Our astrometric and photometric reduction of the data set closely
followed the procedures described in detail in
\citet{2014ApJ...797..115B} and in \citet{2017ApJ...842....6B},
respectively. Below we provide a brief, general description of the
entire reduction process, and emphasize the few important differences
and fine tuning that we had applied to the \citet{2014ApJ...797..115B}
and \citet{2017ApJ...842....6B} procedures. We refer the interested
reader to the original publications for an in-depth data-reduction
description.

\subsection{First-pass photometry}\label{ss:2.1}

All ACS/WFC and WFC3/UVIS \texttt{\_flt}\footnote{\texttt{\_flt}
  exposures are dark and bias corrected and have been flat-fielded,
  but no resampling is applied. Our photometry and astrometry is based
  on \texttt{\_flt} images because they preserve the un-resampled
  pixel data for stellar-profile fitting.} exposures were
pipeline-corrected to minimize the loss of charge-transfer efficiency
(CTE), by means of the empirical pixel-based CTE correction described
in \citet{2010PASP..122.1035A}. The WFC3/IR detector is based on a
different read-out architecture, and it is not affected by CTE
losses.

Next, we derived spatially-variable perturbation PSF models for each
exposure, based on the few hundreds of bright, isolated, unsaturated
stars within. The perturbation models were then combined to the
spatially-variable---but time constant---empirical PSF libraries
(e.g., \citealt{ak06}) to account for telescope breathing effects
(\citealt{dinino08}). These image-tailored PSF models were then used
to measure stellar positions and fluxes in each exposure using the
\texttt{FORTRAN} code \texttt{hst1pass}, which is an advanced version
of the family of camera-dependent \textit{HST} codes based on the
\texttt{img2xym\_WFC} software package
(\citealt{ak06}).\footnote{\url{http://www.stsci.edu/~jayander/CODE/}}
The routine \texttt{hst1pass} run a single pass of source finding for
each exposure, and does not perform neighbor subtraction. The code
then applies the appropriate camera-dependent reduction
routines. Stellar positions in each single-exposure catalog
(thereafter, first-pass catalog) were corrected using the state-of-the
art geometric-distortion corrections of \citet[ACS/WFC]{ak06},
\citet[WFC3/UVIS]{bb09, b11}, and the
publicly-available WFC3/IR correction developed by
J.\ Anderson.\footnote{\url{http://www.stsci.edu/~jayander/STDGDCs/}}

\subsection{The master frame}\label{ss:2.2}

We cross-matched unsaturated stars in our first-pass catalogs with
those in the Gaia data release 1 (DR1, \citealt{2016A&A...595A...2G,
  2016A&A...595A...1G})\footnote{\url{http://gea.esac.esa.int/archive/}}
within 3 arcmin from the center of field F1. We found 131 sources in
common, which were used as a reference to:\ (1) register our stellar
positions to the Gaia-DR1 absolute astrometric system\footnote{Note
  that Gaia DR1 positions refer to epoch 2015.0 and are given with
  respect to the Internatioanl Celestial Reference System, ICRS.}; and
(2) define a right-handed, pixel-based, Cartesian reference frame (the
master frame) with the X and Y axes parallel to the R.A. and
Dec. directions, respectively, and with a pixel scale of exactly 40
\maspx (nearly the same as that of the WFC3/UVIS).

Then, we applied general, six-parameter linear transformations to
transform the stellar positions of each first-pass catalog into the master
frame, with which we created preliminary epoch- and filter-dependent
average catalogs of positions and fluxes.  The instrumental average
photometry of these preliminary catalogs was obtained by rescaling the
fluxes of each exposure to the instrumental magnitude zero-point of
the first long exposure taken in each filter/epoch.  The linear
transformations we applied are based on likely \wcen\ members on the
basis of their positions on the instrumental F606W versus
F606W$-$F814W CMD, so as to minimize large positional residuals due to
stellar PMs of field stars over $\sim 15.1$ years.

\subsection{Second-pass photometry}\label{ss:2.3}

The \texttt{FORTRAN} software package KS2 (Anderson in preparation,
see \citealt{2017ApJ...842....6B} for details) allows us to
simultaneously measure stars in all the individual exposures and for
the entire set of filters. KS2 is the evolution of
\textit{kitchen\_sync}, originally designed to reduce specific ACS/WFC
data (\citealt{2008AJ....135.2055A}). KS2 takes the results of the
first-pass photometry and the information coming from the
six-parameter transformations into the master frame, and uses all the
exposures together to find, measure, and subtract stars in several
waves of finding, moving progressively from the brightest to the
faintest stars. We closely followed most of the reduction
prescriptions given in \citet{2017ApJ...842....6B}, but we also
applied a few critical changes (described below) aimed at
maximizing the astrometric precision of the reduction.

KS2 is designed to work best in moderately crowded fields, so that it
is limited by design to a 2-pixel search radius around the peak
defined by each source on the master frame. Because we expect a
significant fraction of field objects to have moved by more than 2
pixels (80 mas) over $\sim 15.1$ years with respect to the mean motion
of the cluster, we decided to run KS2 over two distinct sets of
data:\ GO-9444 and GO-10101 data (taken in 2002 and in 2005), GO-14118
and GO-14662 data (taken in 2015 and 2017). Both runs have long F606W
and F814W exposures\footnote{Note that the transmission curves of the
  these two filters for the ACS/WFC and the WFC3/UVIS detectors are
  similar but not identical.}, so the finding
stage of KS2 is based on these. This choice maximizes the number of
common sources found in the two runs. Both KS2 runs performed nine
finding passes, which allowed us to find and measure stars down to
near the cluster's hydrogen-burning limit (HBL).

KS2 measures stars in three different ways. The first method (method
1) applies when a star is bright enough to generate a distinct peak
within its central $5\times 5$-pixel, neighbor-subtracted raster.  In
this case, KS2 measures the position and the flux of this star using
the appropriate PSF model for the star's location in an exposure. The
other two methods do not fit stellar positions in each exposure, but
rely on the positions determined during the finding stage. Although
method 1 does not allow us to obtain photometry of stars as faint as
those recovered by methods 2 and 3, method 1 is the only method for
which stellar positions are solved for in each individual exposure. As
such, for the remainder of this paper, we will consider only stellar
positions as measured by the KS2 method 1.

In addition to the photometric diagnostics described in
\citet{2017ApJ...842....6B}, we also had KS2 output the \texttt{RADXS}
parameter (\citealt{2008ApJ...678.1279B}). \texttt{RADXS} tells us how
much flux a source has with respect to the PSF predictions just
outside the PSF core.  Galaxies and blends have large positive values
of \texttt{RADXS}, while objects sharper than the PSF, e.g. cosmic
rays or hot pixels, have large negative values of \texttt{RADXS}.

KS2 found a total of 28\,345 sources in the first run based on ACS/WFC
exposures, and 14\,883 sources in the second run based on WFC3
exposures, due to the smaller field of view (FoV) of the latter data
set. In both cases, sources are measured in both F606W and F814W
filters.  We cross-identified the objects in these two lists and found
8751 sources measured in all four epochs. These sources constitute the
master list that we will use in Sect.~\ref{ss:2.5} to compute PMs.

\subsection{Photometric calibration}\label{ss:2.4}

Instrumental magnitudes were zero-pointed to the Vega-mag flight
system following the prescriptions given in
\citet{2017ApJ...842....6B}. We employed a radius of $0\farcs4$ (10
pixels on the master frame) for the aperture photometry on the
\textit{HST}-pipeline-calibrated \texttt{\_drc} and \texttt{\_drz}
images. The adopted ACS/WFC Vega-mag zero points are from
\citet{2016AJ....152...60B}. For the WFC3 filters, Vega-mag zero
points are available at the official WFC3 zero-point
website\footnote{\url{http://www.stsci.edu/hst/wfc3/phot_zp_lbn}}.

Panels (b) and (c) of Fig.~\ref{fig:1} show the visual $m_{\rm F606W}$
versus $m_{\rm F606W}-m_{\rm F814W}$ and the IR $m_{\rm F110W}$ versus
$m_{\rm F110W}-m_{\rm F160W}$ CMDs of the 8751 sources measured in all
four epochs. Saturation in the long F606W exposures starts at $m_{\rm
  F606W}\approx 20.4$. Photometry of brighter stars is based only on
the short exposures (four per filter) of GO-14118 and GO-14662.

\subsection{Proper-motion measurements}\label{ss:2.5}

As shown in \citet{b11}, filters bluer than F336W are not suitable for
high-precision astrometry, because of the presence of color-dependent
residuals in the UVIS distortion solution. Therefore, F275W exposures
were not used to compute high-precision PMs. 

Unlike the case of WFC3/UVIS and ACS/WFC detectors
(\citealt{2014ApJ...797..115B}), the characterization of systematic
effects possibly affecting astrometry in the WFC3/IR has just begun
(e.g., \citealt{2017AJ....153..243Z}).  In addition, the WFC3/IR
detector suffers from a more severe undersampling of the PSF and has a
significantly larger pixel scale ($0\farcs 13$\,pixel$^{-1}$) with
respect to both the WFC3/UVIS ($0\farcs 04$\,pixel$^{-1}$) and ACS/WFC
($0\farcs 05$\,pixel$^{-1}$). We initially included WFC3/IR exposures
to compute PMs, but we found negligible improvements in terms of PM
errors and number of measured sources with respect to using only WFC
and UVIS exposures. Therefore, for this particular study, IR exposures
were not included in the analysis.

Proper motions are computed by closely following the procedures
described in detail in \citet{2014ApJ...797..115B}, in which each
individual exposure is treated as a stand-alone epoch, and PMs are
iteratively obtained as the slope of the straight-line fits to the
master-frame stellar positions versus epoch of observation. The main
process is divided into five steps:\ (1) measure stellar positions in
each exposure; (2) define/improve a common master list based on a set
of reference stars at a specific epoch; (3) cross-identify stars
measured in different exposures with those in the master list; (4)
transform the (X,Y) position of each star as measured in all the
exposures where the star is found on to the master frame, using a local
network of reference stars; (5) fit straight lines to the master-frame
transformed positions versus epoch. The slope of the fit provides a
direct measurement of the PM.

We already have all the necessary pieces of information for step (1).
Unlike \citet{2014ApJ...797..115B}, which made use of stellar
positions as measured by the first-pass photometry, here we started
from the stellar positions measured by KS2 method 1. KS2 offers a
significant advantage over the first-pass photometry, since it
deblends each source prior to fitting of the PSF.  Steps (2), (3),
(4), and (5) are nested into each other, and each of them is iterated
in order to:\ (i) allow for the rejection of discrepant observations;
and (ii) improve the reference star list, the master-frame
transformations, and the PM measurements themselves.

As a common reference frame we started with the Gaia-based master
frame defined in Sect.~\ref{ss:2.2}. We also defined an initial set of
likely cluster members (the reference stars) extracted from the master
list on the basis of their location on the CMD.  Steps (2) to (5) are
iterated as detailed in \citet{2014ApJ...797..115B}. At the end of
each iteration, the list of reference stars is improved by removing
all objects for which the PM is not consistent with the cluster's mean
motion.

\begin{figure*}[th!]
\centering
\includegraphics[trim=40 165 40 360,clip=true,width=\textwidth]{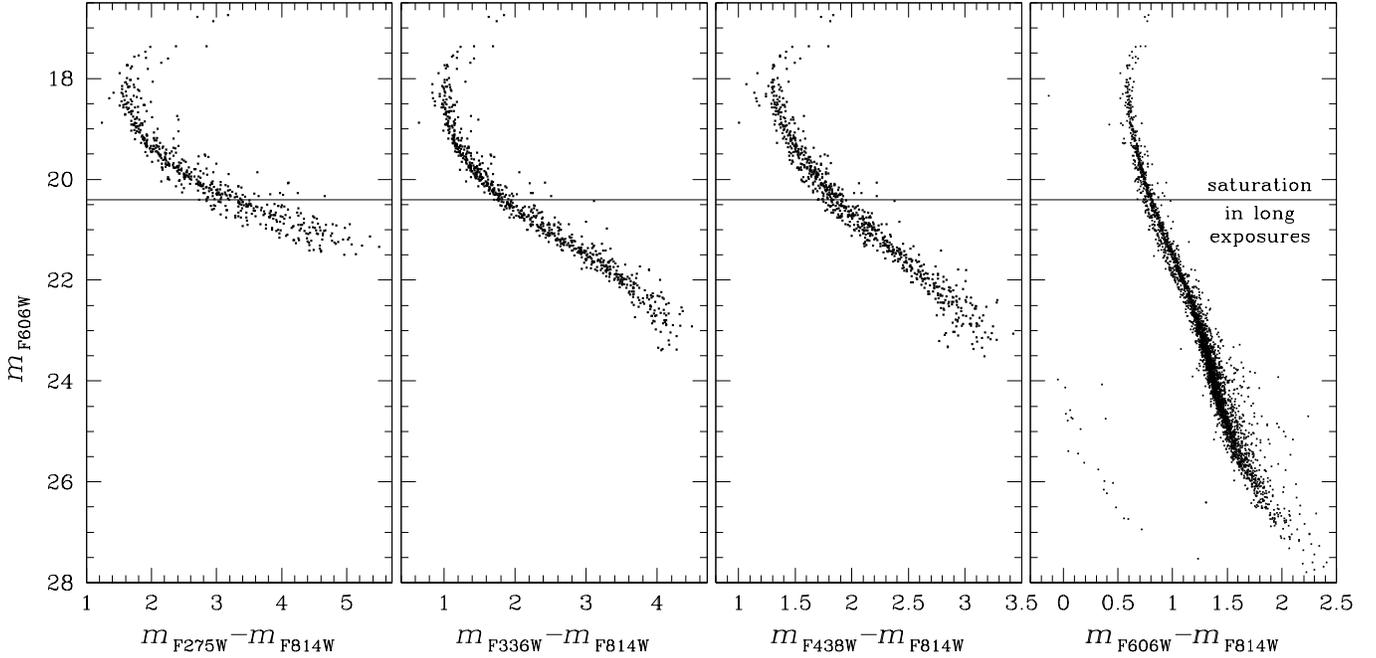}
\caption{From left to right, $m_{\rm F606W}$ versus $F-m_{\rm F814W}$
  CMDs, with $F$ increasing from $m_{\rm F275W}$ to $m_{\rm F606W}$,
  respectively. Each CMD shows only PM-selected cluster members. The
  horizontal line represents the saturation level in long GO-14118 and
  GO-14662 exposures. Stars brighter than this line have PMs obtained
  over a temporal baseline of just two years. Fainter stars have PMs
  computed over temporal baselines of at least 11.7 years.}
\label{fig:2}
\end{figure*}

For each star $i$ in each exposure, position transformations on to the
master frame are typically based on the subset of reference stars
(typically a few hundreds) that are within the same detector amplifier
of the star $i$ (to minimize the impact of uncorrected
geometric-distortion and CTE-mitigation residuals). Only at the last
iteration we further restrict the subset of reference stars to the
closest 45 to the star $i$ (the so-called local transformations, e.g.,
\citealt{2006A&A...454.1029A}).

Proper-motion fitting and data rejection are performed exactly as
described in \citet{2014ApJ...797..115B}. Briefly, for each star we
fit all the transformed X and Y positions coming from different
exposures as a function of the exposure epoch by means of a
least-squares straight line.  After obvious outliers have been
rejected, the residuals of the fit of both X and Y positions are
collected together and rescaled so that, to the lowest order, their
distribution should be consistent with a two-dimensional Gaussian. We
iteratively rejected one data point at a time if its combined
residuals are consistent with a two-dimensional Gaussian distribution
to less than 2.5\% confidence level (see Sect.~5.5 of
\citealt{2014ApJ...797..115B} for more details).

Because of the large internal velocity dispersion of \wcen\ in field
F1 ($\sim 0.34$ \masyr), cluster stars have moved on average by about
0.13 pixels between 2002 and 2017. Field objects have moved on average
much more than that, $\gtrsim 2.5$ pixels, forcing us to allow for a
generous search radius (5 pixels) to cross-identify stars of each
exposure with the master list.  This inevitably led to some
misidentifications.  At the end of the first iteration, PMs can be
used to estimate the master list positions at the precise epoch of
each exposure. This allowed us to adopt a much tighter search radius
(0.75 pixels) to minimize the inclusion of false positives.  Moreover,
we redefined the master list so that the position of its sources is
referred to the average epoch of the data (the reference
epoch):\ 2010.42285.\footnote{\citet{2014ApJ...797..115B} used instead
  GO-10775 (PI:\ Sarajedini, A.) average epochs as the reference epoch
  for each of their analyzed clusters.}  We iterated steps (2) to (5)
a few more times, until the predicted master-list positions at the
reference epoch changed by less than 0.001 pixel and the number of
reference stars remains constant from one iteration to the next.

The initial master list contained 8751 sources, but we were able to
compute high-precision PMs for only 5153 sources. The missing 3598
sources were rejected during the various PM-measurement iterations
according to a variety of different reasons, all related to data
quality and self consistency. All rejection criteria are listed and
described in \citet{2014ApJ...797..115B}.  Our final PM catalog is
supplied with the same set of quality and diagnostic parameters
described in \citet{2014ApJ...797..115B}. 

Panel (d) of Fig.~\ref{fig:1} shows the PM diagram of the 5153
measured sources. Because our reference list consists of cluster
members, our PMs are \textit{relative} to the bulk motion of the
cluster, and cluster members are gathered together around (0,0). The
lesser clump of sources at about (+3.5,+6.5) is mostly populated by
background galaxies, and will be the subject of the next paper in this
series (\citealt{libra.p3}). All other sources are foreground and
background field stars. Note that we refer to \dmacd\ and \dmd\ PM
components, rather than $\mu_\alpha \cos \delta$ and $\mu_\delta$,
because we want to emphasize the fact that our PMs are relative to the
cluster's mean motion.

Panel (e) shows the distribution of the temporal baselines used to
compute the PM of each star. Because stars brighter than $m_{\rm
  F606W}\sim 20.4$ are unsaturated only in the (few) short exposures
of GO-14118 and GO-14662, their PM is based on a temporal baseline of
just $\sim 2$ years (yellow). The vast majority of the remaining stars
are measured in all four epochs, providing a temporal baseline of
$\sim 15.1$ years (green). Finally, sources outside the GO-9444 FoV
but measured in the other three epochs have PMs computed over a
temporal baseline of $\sim 11.7$ years (red).  The (X,Y) position of
sources with measured PMs are shown on the pixel-based master frame in
panel (f), color-coded according to the relevant temporal baseline
used to compute their PM. Finally, PM errors as a function of the
$m_{\rm F606W}$ magnitude are in panel (g), also color-coded according
to the temporal baseline used. Clearly, the PM error of bright stars
(yellow) is significantly larger than that of faint stars, which are
based on much longer temporal baselines and larger number of
images. Well-measured stars in the long exposures ($20.4\lesssim
m_{\rm F606W}\lesssim 22.4$) have PM errors of about
10\,$\mu$as\,yr$^{-1}$. (Note that Gaia's expected end-of-mission PM
precision at the faint limit ($\gtrsim 0.9$\,mas\,yr$^{-1}$ at $m_{\rm
  F606W}\sim 21.5$) is over a factor of 90 worse,
\citealt{2017MNRAS.467..412P}).

Exposures taken with the magic trio of filters (F275W, F336W and
F438W) are significantly shallower than those taken with the F606W and
the F814W. This can be clearly seen in Fig.~\ref{fig:2}, in which we
plot four different CMDs of the form $m_{\rm F814W}$ versus
$F-m_{F814W}$, with $F$ varying from F275W to F606W from the left to
the right panels. In all panels, only stars whose motion is within 1.5
\masyr\ of the cluster's mean motion are shown. The F606W saturation
level in the long exposures is marked in all panels. Clearly, F275W
exposures are the least complete.

\begin{figure*}[th!]
\centering \includegraphics[trim=50 310 50
  100,clip=true,width=\textwidth]{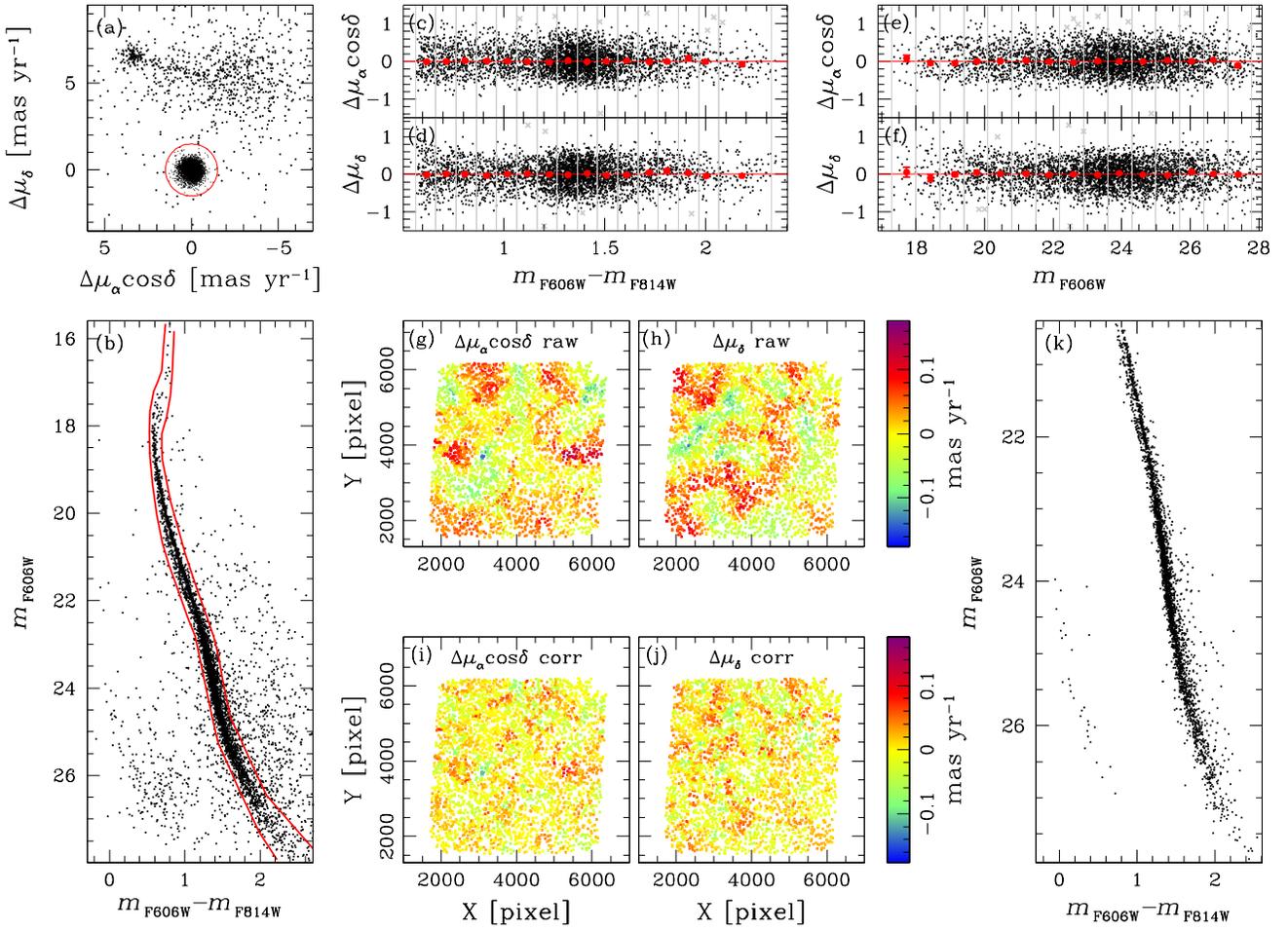}
\caption{This figure illustrates the a-posteriori correction
  procedures applied to the raw PM measurements. (a) PM diagram of the
  sources in field F1. We preliminary select cluster members as all
  objects within the red circle of radius 1.5 \masyr. (b) An
  additional selection makes use of stellar positions on the $m_{\rm
    F606W}$ versus $m_{\rm F606W}-m_{\rm F814W}$ CMD (within the two
  red lines drawn by hand). Panels (c) and (d) show that raw PMs do
  not suffer from systematic effects as a function of stellar
  color. Similarly, panels (e) and (f) show that raw PMs do not suffer
  from systematic effects as a function of stellar magnitude. Note
  that the horizontal red lines at $\Delta\mu=0$ in these four panels
  are not a fit to the data, but indicate lack of trends.  In panels
  (g) and (h) we report the maps of the locally-measured (closest 100
  stars) mean raw PM components of cluster members. Specifically, the
  deviation along $\mu_\alpha\cos\delta$ is in panel (g), and the
  deviation along $\mu_\delta$ is in panels (g) and (h).  Each star is
  color-coded according as shown by the vertical bar the immediate
  right of panel (h). Panels (i) and (j) show the maps of the
  locally-measured mean corrected PM of cluster members. We applied
  the same color-scheme as in panels (g) and (h). Finally, panel (k)
  shows the $m_{\rm F606W}$ versus $m_{\rm F606W}-m_{\rm F814W}$ CMD
  of cluster members with high-quality PMs. See the text for details.}
\label{fig:3}
\end{figure*}

\subsection{Proper-motion corrections}\label{ss:2.6}

Following the prescriptions given in \citet[their Sects.~7.3 and
  7.4]{2014ApJ...797..115B} we applied a-posteriori corrections to
mitigate any residual source of systematic errors. We started by
selecting likely cluster members on the basis of their position on the
PM diagram (within 1.5 \masyr\ from the bulk distribution, red circle
in panel (a) of Fig.~\ref{fig:3}) and of their position on the $m_{\rm
  F606W}$ versus $m_{\rm F606W}-m_{\rm F814W}$ CMD (within the two red
lines shown in panel (b) of the same figure). 

We verified that neither component of the PM suffers from systematic
effects due to stellar color (panels c and d) and luminosity (panels e
and f). In each of these panels, we divided the sample into equally
populated bins in color or in magnitude (within the gray vertical
lines). Within each bin, we computed the 3$\sigma$-clipped median
value of the motion along \dmacd\ and \dmd. Rejected stars are shown
with gray crosses. The computed median values are shown as red solid
circles, with errorbars. The associated errors of the mean are
typically smaller than the size of red circles. The horizontal red
line is not a fit to the data, but indicates lack of systematic
effects. It is clear from panels (c), (d), (e), and (f) that color and
luminosity systematic effects, if present, are negligible.

We did notice marginal, spatially-varying systematic effects as a
function of stellar positions on the master frame. These spatial
effects are due to small single-exposure CTE and geometric-distortion
residuals that, given the relative small number of different roll
angles employed (up to three), do not cancel out. These effects can be
divided into a low- and a high-frequency variation.  The low-frequency
variation correlates well with the map of the temporal baseline used
to compute PMs, which in turn reflects the number and type of
overlapping images at any given location on the master frame. We
corrected this systematic effect by computing three median values of
each component of the motion of selected cluster members, one for each
of the three groups of temporal baselines shown if Fig.~\ref{fig:1}e,
and subtracting them from the motion of each star according to the
temporal baseline used. By construction, these median values should
all be equal to zero. Instead, we found deviations of the order
of a few tens of $\mu$as\,yr$^{-1}$ (note that the measured cluster
dispersion is much larger, about $\sim 0.34$\,mas\,yr$^{-1}$).

High-frequency-variation systematic effects were corrected as
described in Sect.~7.4 of \citet{2014ApJ...797..115B}. In a brief,
both components of the motion of each star were corrected according to
the median value of the closest 100 likely cluster members (excluding
the target star itself). Again, by construction the median should be
equal to zero, and the measured deviation is used as the local
correction. Note that here local corrections are only based on
distance and not on magnitude, as it was the case in
\citet{2014ApJ...797..115B}, because our PMs do not suffer from
systematic effects due to stellar color or magnitude (panels (c), (d),
(e) and (f) of Fig.~\ref{fig:3}).

Panels (g) and (h) of Fig.~\ref{fig:3} show the maps of the local
median values obtained with the uncorrected (raw) components of the
motion; \dmacd\ in panel (g) and \dmd\ in panel (h). Each point is a
source, color-coded according its locally-averaged PM value, as shown
on the color bar on the right-hand side of panel (g). Panels (i) and
(j) show similar maps after the high-frequency variations are
corrected. Points are color-coded using the same color scheme as
panels (g) and (h). The overall size of the local PM variations is
significantly lower than those of the uncorrected maps. The
uncorrected maps have a root mean square (RMS) of 42
$\mu$as\,yr$^{-1}$. The RMS of the corrected maps is 27
$\mu$as\,yr$^{-1}$ (i.e., about 8\% of the intrinsic velocity
dispersion of the cluster).

Both the low-frequency and the high-frequency a-posteriori corrections
come at a cost. In fact, the corrections we applied to each star are
based on the median value of a sample of cluster members, which have
their own intrinsic dispersion. As a result, there is an error
associated to these corrections, i.e., the standard error of the mean.
For instance, the high-frequency variations were corrected using the
median motion of 100 stars, so that the associated error is of the
order of $\sim 0.34/\sqrt{100}\,{\rm mas}\,{\rm yr}^{-1}\sim 34\,
\mu{\rm as}\,{\rm yr}^{-1}$ per coordinate. Similar (but smaller, due
to the larger sample size) correction errors are found for the
low-frequency variation.  The effects of the increased PM errors are
reflected in a marginally larger (but rounder) cluster dispersion on
the PM diagram.  The total PM errors associated to the a-posteriori
corrected PMs are the sum in quadrature of the raw PM errors and both
the low- and high-frequency correction errors.  Since different
scientific investigations favor the use of one or the other way of
estimating PMs, our final PM catalog contains both the raw and the
corrected PMs, with uncertainties (see Sect.\ref{A1}).

\subsection{Proper-motion selections}\label{ss:2.7}

We aim to study the finest kinematic details of the mPOPs of the
cluster, so we applied a few selection criteria to our PM catalog in
order to analyze the best measured sources. To start, we selected only
stars for which the computed PM is based on the longest available temporal
baseline ($\sim 15.1$\,yr). In addition, we restricted the sample to
stars that are:\ (1) measured in at least 10 distinct exposures; (2)
with a rejection rate of less than 15\%, (3) with reduced $\chi^2<4$
in each coordinate; (4) with fluxes above $3\sigma$ the local sky
background; (5) with \texttt{RADXS} values in F606W and F814W between
$-0.05$ and 0.05; (6) \texttt{QFIT} values in F606W and F814W greater
than 0.7; and (7) with PMs within 1.5 \masyr with respect to the
cluster's bulk motion, a value $\sim 4.5$ times larger than the
observed cluster dispersion. With this initial list, we iteratively
rejected stars with PM errors larger than 50\% of the local velocity
dispersion, as described in Sect.~7.5 of \citet{2014ApJ...797..115B}.

These selection procedures were applied to both the raw and the
corrected PMs, and we run a parallel analysis on both sets of PM
measurements. The resulting velocity-dispersion profiles for the raw
and the corrected PM measurements were consistent with each other well
within the uncertainties. In what follows, we present the analysis on
the kinematics of the mPOPs of the cluster based on corrected PM
measurements. The final list of high-precision, PM-selected cluster
members includes 2912 sources, extending from $m_{\rm F606W}\simeq
20.4$ down to near the HBL at $m_{\rm F606W}\sim 28$. Panel (k) of
Fig.~\ref{fig:3} shows the $m_{\rm F606W}$ versus $m_{\rm
  F606W}-m_{\rm F814W}$ CMD of these 2912 cluster stars.

\begin{figure*}[t!]
\centering
\includegraphics[trim=35 525 45 100,clip=true,width=1.\textwidth]{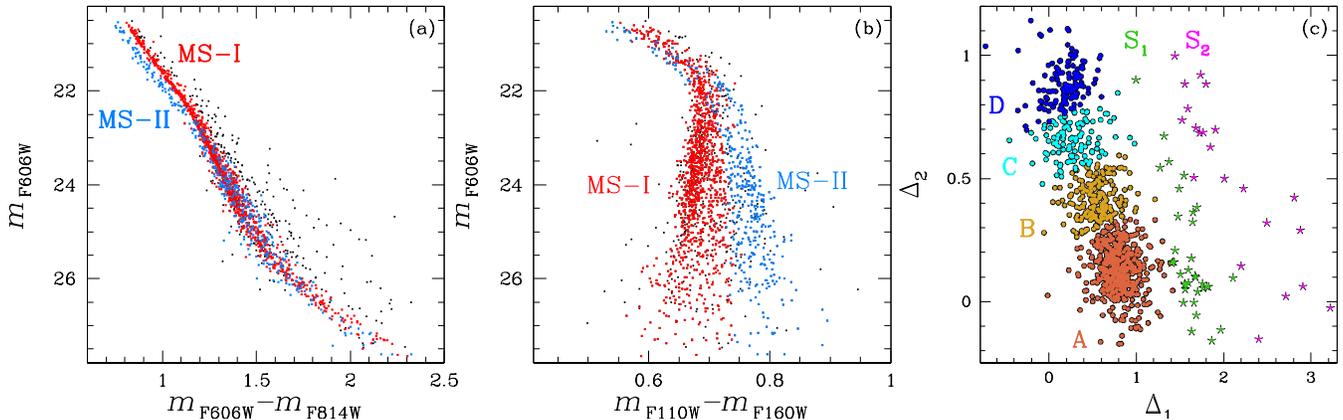}
\caption{(a) MS-I (red) and MS-II (azure) stars on the \mv\ versus
  \mvi\ CMD. (b) MS-I and MS-II stars on the \mv\ versus
  \mjh\ CMD. (c) The $\Delta_2$ versus $\Delta_1$ chromosome map (see
  Paper~I) with the six populations identified below the MS knee. The
  four main populations are:\ A (orange), B (yellow), C (cyan) and D
  (blue). The two lesser subgroups are S$_1$ (green) and S$_2$
  (magenta). All populations are selected as in Paper~I. See the text
  for details.}
\label{fig:4}
\end{figure*}

\section{Multiple-population kinematics}\label{sec:3}

\subsection{Naming conventions}\label{ss:3.0}

The MS of \wcen\ has been known for 20 years to be split into two main
components (e.g., \citealt{anderson97, 2004ApJ...605L.125B}), which
were historically named the blue MS (bMS) and the red MS (rMS), and
a lesser, ``anomalous'' component (MSa) associated with the peak of the
[Fe/H] distribution. Hints of an even more complicated scenario were
discovered by \citet{2010AJ....140..631B}, in which both the rMS and
the MSa turned out to be themselves split into two subcomponents.

More recently, thanks to the same five UVIS filters used in this large
program, distinct mPOPs have been detected in all the GCs analyzed so
far, including \wcen\ (see, e.g., \citealt{2015AJ....149...91P, 
  2017ApJ...844..164B, 2017MNRAS.464.3636M}, and references
therein). In particular, the latter authors identified up to 17
populations on the RGB, with each of them possibly related to distinct
Fe, O, and Na abundances (\citealt{2011ApJ...731...64M}). At the
bright MS level, \citet{2017ApJ...844..164B} isolated at least 15
distinct MS populations in the cluster core, organized into 5 main
population groups, which they named as:\ rMS, bMS, MSd, MSe, and
MSa. (Please note that the rMS and the bMS defined in
\citealt{2017ApJ...844..164B} are actually subcomponents of the
historical rMS and bMS, respectively.)

In particular, the rMS, the bMS, and the MSd groups as defined in
\citet{2017ApJ...844..164B} are each split into three subcomponents
(e.g., the rMS is made up of rMS1, rMS2, and rMS3 subpopulations); the
MSe is made up of four subcomponents, while MSa is divided into two
subcomponents. Both the bMS and the MSd groups are consistent with
stellar populations being highly enriched in He and moderately
enriched in Fe with respect to the rMS group. The MSe group, on the
other hand, shares similar properties to the rMS. Finally, both MSa
subpopulations are likely highly enriched in both He and Fe.

Since both the historical rMS and bMS populations are actually made up
of several population subgroups, in Paper~I we decided to use the
labels MS-I and MS-II, respectively, to refer to these two main MS
branches that can be seen on a visual CMD.  In addition, in Paper~I we
also made use of the IR filters F110W and F160W to identify for the
first time four main groups of stars below the MS knee, which we named
as populations A, B, C, and D, plus two lesser MS components that we
called populations S$_1$ and S$_2$. These populations merge together
in the proximity of the MS knee, and above the knee only the MS-I and
MS-II meta-groups can be identified using IR filters and/or visual
filters.

Clearly, there is quite some room for confusion with population
names. To complicate things, different authors have used different
names to refer to same populations of \wcen, especially along the
RGB.  In what follows, we will adopt the meta-group naming convention
of Paper~I. We will also use the terms first-generation
(1G) and second-generation (2G) stars to refer to the stars of the
meta-groups MS-I and MS-II, respectively, to emphasize differences in
their dynamical-evolution histories. 1G stars in \wcen\ are
characterized by primordial (or at most slightly-enhanced) He
abundance, and are generally metal poor. 2G stars are highly He
enhanced and are metal rich. Clear hints of dynamical differences
between 1G and 2G stars in the cluster have been found by studying
their radial distribution (\citealt{2007ApJ...654..915S,
  2009A&A...507.1393B}), with 2G significantly more centrally
concentrated than 1G stars.

Finally, we will use the same labels of Paper~I for the six population
groups identified below the MS knee using IR photometry, and the same
names of \citet{2017ApJ...844..164B} for the five population groups
identified on the bright MS using UV photometry.

\subsection{Subpopulation selections}\label{ss:3.1}

As we can see in Fig.~\ref{fig:2}, the photometry obtained with the
magic trio of filters in our field F1 is dramatically shallow
(especially in F275W), to the point where no meaningful kinematic
properties can be derived if we use them to isolate the different
mPOPs of the cluster, as done in
\citet{2017ApJ...844..164B}.\footnote{Although our UV photometry is as
  shallow as that of \citet{2017ApJ...844..164B}, they were able to
  use it for their mPOPs selections because of the availability of over
  a factor of four more images and the presence of over a factor of
  $\sim 100$ more stars in the central cluster field than in our outer
  field F1.}  Therefore, we applied the same selection criteria of
Paper~I (as well as the high-quality photometry of Paper~I) to isolate
stars in the MS-I and MS-II meta-groups, as well as in the six
populations below the MS knee.

Panels (a) and (b) of Fig.~\ref{fig:4} show MS-I (1135 stars, red) and
MS-II stars (385 stars, azure), as defined in Paper~I, on the
\mv\ versus \mvi\ and on the \mv\ versus \mjh\ CMDs, respectively. The
$\Delta_2$ versus $\Delta_1$ chromosome map of stars below the MS knee
is in panel (c), with the six populations identified and color-coded
as in Paper~I:\ A (523 stars, orange), B (202 stars, yellow), C (117
stars, cyan), D (133 stars, blue), S$_1$ (36 stars, green), and S$_2$
(25 stars, magenta). We refer the reader to Paper~I for a detailed
description of the population selections. All stars plotted in
Fig~\ref{fig:4} have high-precision PM measurements.

\begin{figure*}[t!]
\centering
\includegraphics[trim=25 155 20 80,clip=true,width=\textwidth]{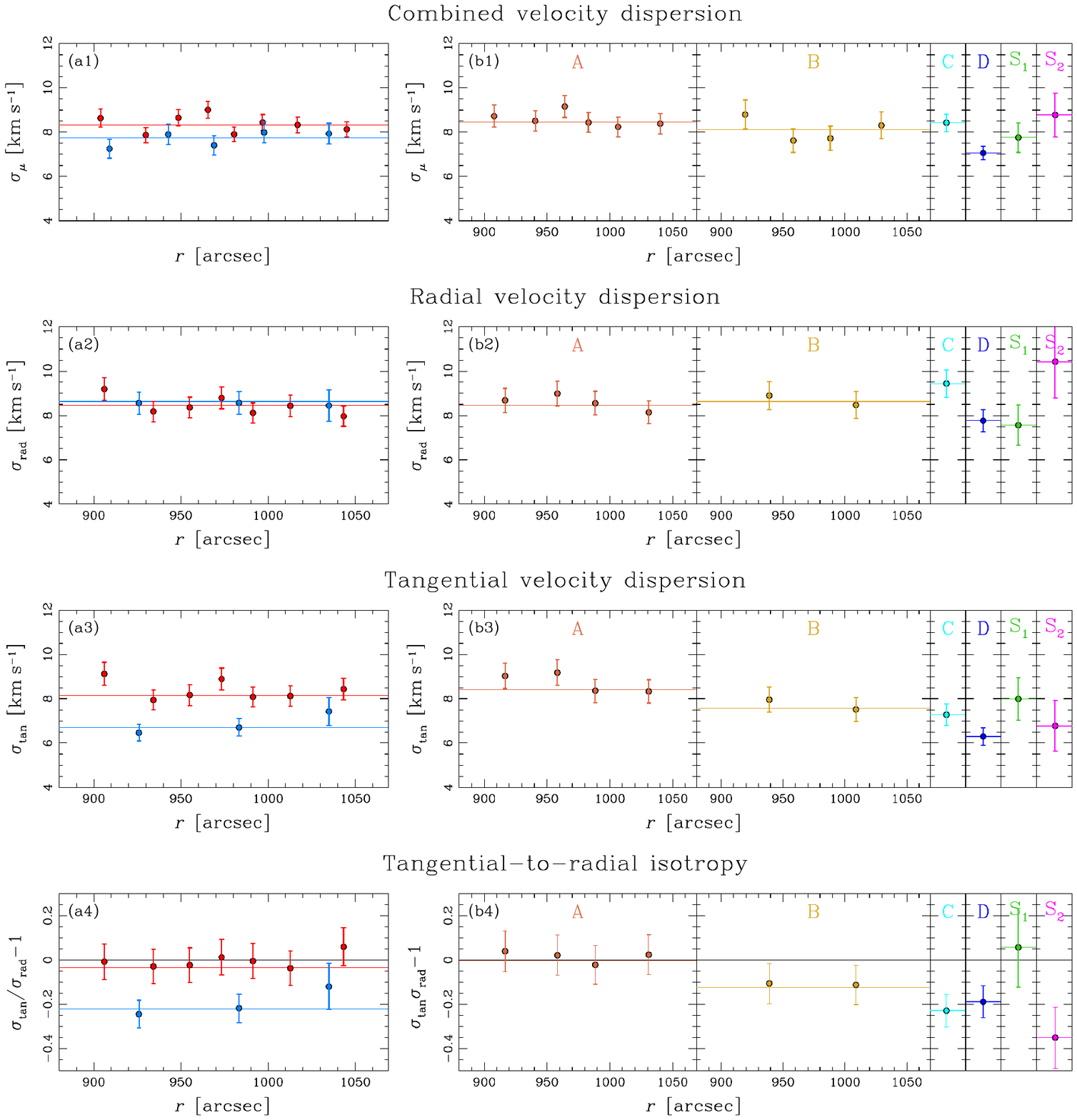}
\caption{(a1) velocity dispersion $\sigma_\mu$ as a function of the
  distance from the cluster's center, $r$, for MS-I (red) and MS-II
  (azure) stars. (b1) $\sigma_\mu$ versus $r$ for the six populations
  below the MS knee. From left to right:\ A, B, C, D, S$_1$, and
  S$_2$, respectively. Populations are color-coded as in
  Fig.~\ref{fig:4}. Because of low-number statistics, the combined
  velocity dispersion of populations C, D, S$_1$, and S$_2$ is
  computed over the entire FoV. (a2) and (b2) are similar to (a1) and
  (b1) but for \srad, the radial component of the velocity
  dispersion. (a3) and (b3) are for the tangential component of the
  velocity dispersion \stan. (a4) and (b4) show the deviation from
  tangential-to-radial isotropy.  See the text for details.}
\label{fig:5}
\end{figure*}

\subsection{Velocity-dispersion profiles}\label{ss:3.2}

Velocity dispersions are estimated using the same method described in
\citet{2010ApJ...710.1063V}, which corrects the observed scatter for
the individual stellar PM uncertainties. Unless stated otherwise, we
indicate with \sm\ the average one-dimensional velocity dispersion of
the combined \dmacd\ and \dmd\ PM components.  We do not expect
significant differences in the velocity dispersion as a function of
distance from the cluster center within the field F1, because of the
relatively small field size compared to its distance from the cluster
center. Nevertheless, for those populations with at least 200 stars,
namely:\ MS-I, MS-II, A, and B, we measured \sm\ in equally-populated
radial intervals. For the remaining populations C, D, S$_1$, and S$_2$
we derived a single value of the velocity dispersion over the entire
field.  Velocity dispersions are given in units of \kms\ by assuming a
cluster distance of 5.2 kpc (\citealt{h96}).

Panel (a1) of Fig.~\ref{fig:5} shows the velocity-dispersion profiles
\sm\ of the MS-I (red) and MS-II (azure) meta-groups as a function of
the distance from the cluster center. The horizontal red (MS-I) and
azure (MS-II) lines mark the average \sm\ value computed over the
entire field, and are not the average of the \sm\ values of the radial
bins.  Both meta-groups show a flat distribution of \sm\ versus
radius. MS-II (2G) stars appear to be slightly kinematically colder
than MS-I (1G) stars.  The average velocity dispersion of both MS-I
and MS-II stars, about 8\,\kms, is more than a factor of two less than
in the central field (\citealt{2010ApJ...710.1032A}), as a direct
consequence of hydrostatic equilibrium (see also
\citealt{2014ApJ...797..115B}).

Panel (b1) of Fig.~\ref{fig:5} shows the velocity-dispersion profiles
for the six populations below the MS knee. For clarity, profiles of
different populations are shown separately, from population A to
population S$_2$ from the left to the right, respectively. The average
\sm\ values of the populations computed over the entire field (colored
horizontal lines) are all consistent with each other within the
errors, except for population D. We find population D to be
significantly kinematically colder than population A at the 3$\sigma$
confidence level. It is worth to stress that no radial dependence of the
velocity dispersion is present in the analyzed field.

Panels (a2) and (b2) are similar to panels (a1) and (b1) but for the
velocity dispersion of the radial (i.e., towards the center of the
cluster, not along the line of sight) component of the motion
\srad. We find MS-I and MS-II to have the same radial velocity
dispersion. Similarly, we find all of the six populations in (b2) to
have the same radial velocity dispersion within the errors. In panels
(a3) and (b3) we report the velocity-dispersion profiles of the
tangential component of the motion (\stan) for each population. The
\stan\ of the MS-II meta-group is significantly lower, at the
3.2$\sigma$ level, than that of the MS-I. Below the MS knee we find
the average \stan\ of populations B, C, and D to be progressively
lower (and with increasing confidence level) than that of population
A.

The average values of \sm, \srad, and \stan\ for each population, with
associated errors, are listed in Table~\ref{tab:2}.

\begin{table*}[t!]
\caption{Multiple-population kinematics in field F1.\label{tab:2}}
\centering
\scriptsize{
\begin{tabular}{lr|rrrr|rrrr}
\hline\hline
$\!\!\!\!$Population$\!\!\!\!$&$N_{\rm stars}$&\multicolumn{1}{c}{$\sigma_\mu$}&\multicolumn{1}{c}{$\sigma_{\rm rad}$}&\multicolumn{1}{c}{$\sigma_{\rm tan}$}&\multicolumn{1}{c|}{$\sigma_{\rm tan}/\sigma_{\rm rad}-1$}&\multicolumn{1}{c}{$\langle \mu_\alpha \cos\delta\rangle$}&\multicolumn{1}{c}{$\langle \mu_\delta\rangle$}&\multicolumn{1}{c}{$\langle \mu_{\rm rad}\rangle$}&\multicolumn{1}{c}{$\langle \mu_{\rm tan}\rangle$}\\
  $\phantom{o}$        &      $\phantom{o}$       & \multicolumn{1}{c}{(\kms)}    & \multicolumn{1}{c}{(\kms)}          &          \multicolumn{1}{c}{(\kms)}&        $\phantom{o}$      & \multicolumn{1}{c}{(\masyr)}       & \multicolumn{1}{c}{(\masyr)}     & \multicolumn{1}{c}{(\masyr)}&\multicolumn{1}{c}{(\masyr)} \\
\hline
MS-I  $\!\!$&$\!\!$ 1135 $\!\!$&$\!\!$ $ 8.31 \pm 0.13$ $\!\!$&$\!\!$ $ 8.46 \pm 0.19$ $\!\!$&$\!\!$ $ 8.16 \pm 0.18$ $\!\!$&$\!\!$ $ -0.035 \pm 0.031$$\!\!$&$\!\!$ $-0.003\pm 0.010$  $\!\!$&$\!\!$  $0.033\pm 0.011$  $\!\!$&$\!\!$  $-0.022\pm0.010$$\!\!$&$\!\!$ $+0.024\pm0.010$\\
MS-II $\!\!$&$\!\!$  385 $\!\!$&$\!\!$ $ 7.73 \pm 0.22$ $\!\!$&$\!\!$ $ 8.63 \pm 0.34$ $\!\!$&$\!\!$ $ 6.71 \pm 0.27$ $\!\!$&$\!\!$ $ -0.222 \pm 0.044$$\!\!$&$\!\!$ $0.031\pm  0.016$  $\!\!$&$\!\!$  $-0.040\pm0.016$  $\!\!$&$\!\!$  $0.022\pm 0.018$$\!\!$&$\!\!$ $-0.064\pm 0.014$\\
\hline
A     $\!\!$&$\!\!$  523 $\!\!$&$\!\!$ $ 8.44 \pm 0.15$ $\!\!$&$\!\!$ $ 8.45 \pm 0.22$ $\!\!$&$\!\!$ $ 8.43 \pm 0.22$ $\!\!$&$\!\!$ $ -0.002 \pm 0.036$$\!\!$&$\!\!$   $0.011\pm 0.015$ $\!\!$&$\!\!$  $0.043\pm 0.016$  $\!\!$&$\!\!$  $-0.040\pm0.015$$\!\!$&$\!\!$$+0.024\pm0.016$\\
B     $\!\!$&$\!\!$  202 $\!\!$&$\!\!$ $ 8.12 \pm 0.27$ $\!\!$&$\!\!$ $ 8.64 \pm 0.40$ $\!\!$&$\!\!$ $ 7.57 \pm 0.35$ $\!\!$&$\!\!$ $ -0.125 \pm 0.057$$\!\!$&$\!\!$   $-0.010\pm0.023$ $\!\!$&$\!\!$  $0.029\pm 0.024$  $\!\!$&$\!\!$  $-0.018\pm0.025$$\!\!$&$\!\!$$+0.010\pm0.022$\\
C     $\!\!$&$\!\!$  117 $\!\!$&$\!\!$ $ 8.43 \pm 0.40$ $\!\!$&$\!\!$ $ 9.44 \pm 0.63$ $\!\!$&$\!\!$ $ 7.28 \pm 0.49$ $\!\!$&$\!\!$ $ -0.229 \pm 0.073$$\!\!$&$\!\!$   $0.018\pm 0.030$ $\!\!$&$\!\!$  $-0.068\pm0.033$  $\!\!$&$\!\!$  $0.085\pm 0.035$$\!\!$&$\!\!$$-0.108\pm 0.027$\\
D     $\!\!$&$\!\!$  133 $\!\!$&$\!\!$ $ 7.06 \pm 0.31$ $\!\!$&$\!\!$ $ 7.77 \pm 0.49$ $\!\!$&$\!\!$ $ 6.30 \pm 0.40$ $\!\!$&$\!\!$ $ -0.189 \pm 0.072$$\!\!$&$\!\!$   $0.059\pm 0.026$ $\!\!$&$\!\!$  $-0.016\pm0.024$  $\!\!$&$\!\!$  $-0.029\pm0.027$$\!\!$&$\!\!$$-0.033\pm 0.022$\\
S$_1$ $\!\!$&$\!\!$   36 $\!\!$&$\!\!$ $ 7.76 \pm 0.66$ $\!\!$&$\!\!$ $ 7.56 \pm 0.91$ $\!\!$&$\!\!$ $ 8.00 \pm 0.96$ $\!\!$&$\!\!$ $  0.057 \pm  0.18$$\!\!$&$\!\!$   $-0.008\pm0.048$ $\!\!$&$\!\!$  $0.103\pm 0.052$  $\!\!$&$\!\!$  $-0.065\pm0.051$$\!\!$&$\!\!$$+0.173\pm0.049$\\
S$_2$ $\!\!$&$\!\!$   25 $\!\!$&$\!\!$ $ 8.77 \pm 0.99$ $\!\!$&$\!\!$ $10.44 \pm  1.6$ $\!\!$&$\!\!$ $ 6.78 \pm  1.1$ $\!\!$&$\!\!$ $ -0.35  \pm  0.15$$\!\!$&$\!\!$   $0.154\pm 0.087$ $\!\!$&$\!\!$  $0.029\pm0.067$  $\!\!$&$\!\!$  $-0.189\pm0.091$$\!\!$&$\!\!$$-0.098\pm 0.060$\\
\hline
WDs   $\!\!$&$\!\!$   29 $\!\!$&$\!\!$ $8.39 \pm 0.88$ $\!\!$&$\!\!$ $8.98\pm1.32$     $\!\!$&$\!\!$ $7.78\pm1.17$ $\!\!$&$\!\!$$-0.13\pm 0.18$$\!\!$ &$\!\!$  $0.108\pm0.074$ $\!\!$&$\!\!$   $-0.097\pm0.057$ $\!\!$&\multicolumn{1}{r}{$\!\!$$0.02\pm0.10$$\!\!$}&$\!\!$$-0.201\pm 0.088$\\
\hline
\end{tabular}}
\end{table*}

\subsection{Anisotropy}\label{ss:3.3}

Regardless of the exact mechanism(s) that led to the formation of
mPOPs of stars in GCs (e.g., \citealt{decressin, dercole2008,
  bastian}, but see \citealt{2015MNRAS.454.4197R} for a critical
review), all models assume 2G (MS-II) stars to form more spatially
concentrated than 1G (MS-I) stars.  The subsequent long-term dynamical
evolution of the cluster will eventually erase any difference in the
spatial distributions of 1G and 2G stars.  $N$-body simulations (e.g.,
\citealt{2015ApJ...810L..13B}) show that 2G stars gradually diffuse
from the cluster's inner regions to the outskirts (which are initially
dominated by 1G stars) preferentially on radial orbits.  The outskirts
of massive-enough clusters have local two-body-relaxation times long
enough that we should still be able to see the fingerprints of the
dynamical evolution of their mPOPs. Indeed, this has been observed in
47~Tuc (\citealt{2013ApJ...771L..15R}) and in NGC~2808
(\citealt{2015ApJ...810L..13B}). Both 47~Tuc ($\sim 1.3 \times 10^6\,
\mathrm{M}_\odot$, \citealt{2017arXiv171010666H}) and NGC~2808 ($\sim
1.4 \times 10^6\, \mathrm{M}_\odot$, \citealt{2011ApJ...742...51B})
are among the most massive GCs of the Milky Way, and the fields
analyzed by \citet{2013ApJ...771L..15R} and
\citet{2015ApJ...810L..13B} are outside $\sim 2\times r_{\rm h}$.

\wcen\ is the most massive GC of the Galaxy, with an estimated mass of
$\sim 4 \times 10^6\, \mathrm{M}_\odot$
(\citealt{2013MNRAS.429.1887D}), and our field F1 is located at $\sim
3\times r_{\rm h}$ from the center. Therefore, we do expect to detect
significant differences in the internal kinematics between 1G and 2G
stars, in particular in their velocity dispersion along the tangential
direction. Indeed, this is exactly what we find (panels (a2), (a3),
(b2), and (b3) for Fig.~\ref{fig:5}).

The deviation from tangential-to-radial isotropy (that is, the ratio
between the velocity dispersions measured along the tangential and the
radial directions minus one, or \stan/\srad$-$1) of the two MS
meta-groups and of the six populations below the MS knee are shown in
panels (a4) and (b4) of Fig.~\ref{fig:5}, respectively. In both
panels, the black horizontal line marks full isotropy.  We find MS-I
stars to be consistent with being isotropic, while MS-II stars are
significantly radially anisotropic (5$\sigma$ confidence level). Below
the MS knee, populations A and S$_1$ are consistent with being fully
isotropic, while populations B, C, D and S$_2$ appear to be radially
anisotropic (at the 2.2$\sigma$, 3.1$\sigma$, 2.6$\sigma$, and
2.3$\sigma$ confidence levels, respectively).

According to the mPOP dynamical evolution models of
\citet{2015ApJ...810L..13B}, the tangential velocity dispersion of 2G
stars evolves towards smaller values than that of 1G stars, in
agreement with our observations. It is the difference in \stan\ that
is responsible for the differences in the anisotropy of 1G and 2G
stars.

\begin{figure*}[th!]
\centering
\includegraphics[trim=25 290 20 100,clip=true,width=\textwidth]{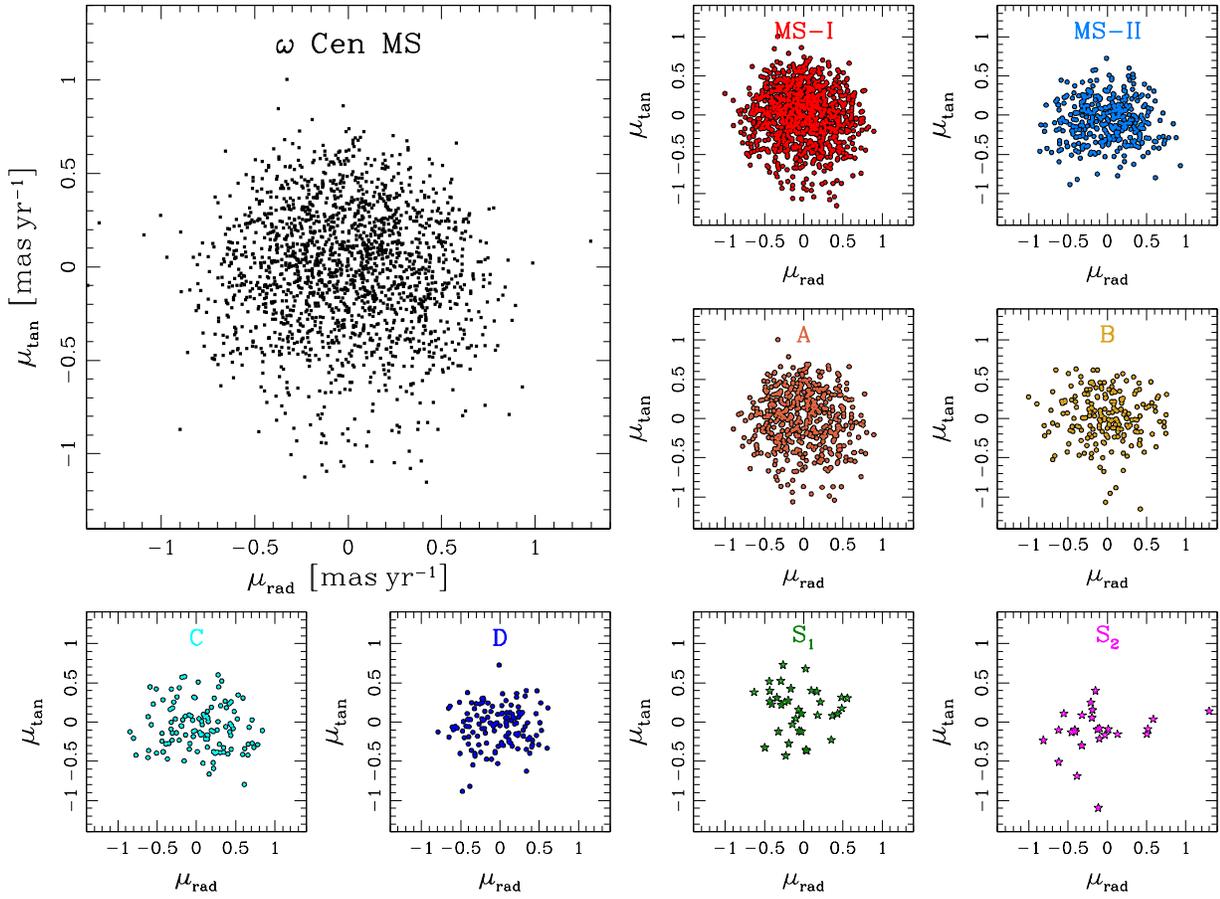}
\caption{The main panel shows the distribution of MS stars in the
  \mtan\ versus \mrad\ plane.  The smaller panels are similar but show
  the individual populations.}
\label{fig:6}
\end{figure*}

\subsection{Differential rotation}\label{ss:3.4}

First of all, we note that all the populations analyzed in this work
have the baricenter of the motion very close to the origin of the PM
diagram.  We find the median value of the motion along the $\alpha
\cos\delta$ direction, $\langle \mu_\alpha \cos \delta \rangle$, of
each population to be consistent with zero with at most marginal
deviations. However, a few small but significant (at the $\gtrsim
3$$\sigma$ level) deviations are present for MS-I, MS-II and
population A stars along the $\delta$ direction. In all three cases,
the median motion along the $\delta$ direction, $\langle \mu_\delta
\rangle$, is of the order of 0.04\,\masyr\ (or just about
1\,\kms). (The $\langle \mu_\alpha \cos \delta \rangle$ and $\langle
\mu_\delta \rangle$ values of each population are reported in
Table~\ref{tab:2}.)  \citet{2002ApJ...573L..95F} combined the
photographic-plate-based PM catalog of \citet{2000A&A...360..472V}
with the RGB photometry of \citet{pancino00}, and found a much larger
difference ($\sim 0.8$\,\masyr) between the relative PM of the RGBa
population (evolved MSa stars) and the cluster's bulk
motion. The culprit of this large PM offset was discovered by
\cite{platais03}. The authors presented a detailed reanalysis of the
\citet{2000A&A...360..472V} PM catalog, and showed that it contains
severe color/magnitude-induced systematic effects. No significant PM
offsets between the different mPOPs of the clusters was later reported
by \citet{2009A&A...493..959B} for RGB stars, and by
\citet{2010ApJ...710.1032A} for bright MS stars in the core. Our PMs
analysis of field F1 populations further support the absence of major
PM deviations between the mPOPs of the cluster.

Analysis of the radial (\mrad) and tangential (\mtan) PM components
provides a more qualitative way to estimate the deviation from
tangential-to-radial isotropy of the mPOPs we found in the previous
Section, as well as a better understanding of the nature of the small
but significant deviation from zero of the relative bulk PM of MS-I,
MS-II, and population A stars. As a reference, we show in the large
panel of Fig.~\ref{fig:6} the PM distribution of all \wcen's MS stars
in the field along the tangential and radial components. The smaller
panels are similar but show the individual populations.  The MS-I and
the population A distributions are visibly rounder than those of the
MS-II and of populations B, C, D, and perhaps also S$_2$, which in
turn are flattened along the \mtan\ direction.  In addition, it
appears that distribution of population S$_1$ is rounder
than---say---the MS-II.

The large panel of Fig.~\ref{fig:6} also reveals that MS distribution
is somewhat skewed towards negative \mtan\ values. A similar behavior
has also been recently reported for stars in an outer field of 47~Tuc
(\citealt{2017arXiv171010666H}). The authors attribute the presence of
the skew to differential rotation of stars in the field.  47~Tuc is
characterized by a high clockwise rotation in the plane of the sky
(see \citealt{2017ApJ...844..167B}), with a peak of the intrinsic
rotational velocity over velocity dispersion $V/\sigma$ of $\sim 0.9$
at around two $r_{\rm h}$, which is where the outer field of
\citet{2017arXiv171010666H} lies. Also the GC \wcen\ is known to be
rotating (counter-clockwise in this case) in the plane of the sky
(e.g., \citealt{2000A&A...360..472V, 2006A&A...445..513V,
  2013MNRAS.436.2598W}), so the presence of skewness in the \mtan\
distribution should not come as a surprise. It is known from studies
of elliptical galaxies that rotation is generally accompanied by
skewness in the line-of-sight velocity distribution
(\citealt{1994MNRAS.269..785B}), and it is therefore natural to expect
this for PMs as well. If we look at the \mtan\ distribution of MS-I
and MS-II stars in Fig.~\ref{fig:6}, it is clear that the skew is
present in the former but not in the latter meta-group. Similarly, a
skew in the \mtan\ distribution is visible for populations A and B,
and perhaps also S$_2$, while the distribution of the remaining
populations seems more symmetric.

The last two columns of Table~\ref{tab:2} list the median values
$\langle$\mrad$\rangle$ and $\langle$\mtan$\rangle$, respectively, for
each population. All $\langle$\mrad$\rangle$ values are consistent
with being zero, with at most marginal deviations (population A). On
the contrary, the behavior of $\langle$\mtan$\rangle$ is
bimodal. Population C and the meta-group MS-II have
statistically-significant (at the $>3$$\sigma$ level) negative
$\langle$\mtan$\rangle$ values. These populations are also
characterized by a lack of skewness in the \mtan\ direction and are
significantly radially anisotropic. Negative but not significant
$\langle$\mtan$\rangle$ values are also found for populations D and
S$_2$, which are also radially anisotropic.  The meta-group MS-I and
populations A, B, and S$_1$ have marginally positive
$\langle$\mtan$\rangle$ values and are also consistent with being
isotropic and having significant skewness (MS-I, A) or even being
tangentially anisotropic (S$_1$). Given that the mPOPs in \wcen\ have
different mean rotation velocities, it is natural that they have
different skewnesses as well.

To quantify the amount of skewness of the \mtan\ distribution and its
statistical significance, we computed two different statistics for
each population:\ (a) the sample skewness values $G_1$ (the normalized
third moment) and the test statistic Z$_{g_1}$ (\citealt{cramer}); and
(b) the third-order Gauss-Hermite moment $h_3$ and its uncertainty
${\rm err}_{h_3}$ (e.g., \citealt{1993ApJ...407..525V}). If $\mu_{\rm
tan}$ has the opposite sign from $G_1$ or $h_3$, then the distribution
has a fatter tail in the direction opposite from the rotation. A
symmetric distribution has $G_1$ and $h_3$ equal to zero. Generally
speaking, a skewness $G_1$ between $-1/2$ and 1/2 indicates an
approximately symmetric distribution. Skew values between $-1$ and
$-1/2$ or between $1/2$ and 1 are indicative of a moderately skewed
distribution.  Highly skewed distributions typically have skewness
$<-1$ or $>1$ (\citealt{bulmer}). These considerations are valid if we
are analyzing data of an entire population, but when the data come
from only a subsample of a population (as it is the case here), then
the subsample can be skewed even though the population is
symmetric. The test statistics Z$_{g_1}$ helps us to quantify the
significance of the skewness of a sample by measuring how many standard
errors separate the sample skewness from zero. If $|{\rm Z}_{g_1}|<2$
(2$\sigma$ confidence level), we can not reach any conclusion about
the skewness of a population, but if $|{\rm Z}_{g_1}|>2$, then the
associated population is likely to be skewed. The statistical
significance of $h_3$ can be assessed directly by its associated
uncertainty. The line-of-sight velocity distributions of elliptical
galaxies typically have $|h_3| \lesssim 0.15$ (e.g.,
\citealt{1993ApJ...407..525V}).

The computed values for $G_1$ and $h_3$ are listed in
Table~\ref{tab:3}. Our results show that there is indeed a significant
skewness in the \mtan\ distribution of the meta-group MS-I and
population A. The skewness has the opposite sign of
$\langle$\mtan$\rangle$, as generally observed in elliptical galaxies
(\citealt{1994MNRAS.269..785B}). The populations B and S$_2$ show
significant evidence of skewness only in $G_1$, but not in $h_3$. The
other populations are consistent with having a symmetric distribution
of \mtan.

\begin{table}[t!]
\caption{Skewness and third Gauss-Hermite values of the
  \mtan\ distribution of each population in field F1.\label{tab:3}}
\centering
\begin{tabular}{lrrrr}
\hline\hline
Population&$G_1$ & Z$_{g_1}$& $h_3$& err$_{h_3}$\\
\hline
MS-I  & $-$0.35    &  $-$4.86 & $-0.074$ & 0.025\\
MS-II & $-$0.03    &  $-$0.20 & 0.016    & 0.044\\
\hline
A     & $-$0.29    &  $-$2.72 & $-0.070$ & 0.029\\
B     & $-$0.62    &  $-$3.61 & 0.011    & 0.051\\
C     &$0.26$      &   $1.17$ & 0.126    & 0.072\\
D     & $-$0.29    & $-$1.40  & $-0.03$  & 0.069\\
S$_1$ & $-$0.24    & $-$0.62  & $-0.21$  & 0.13\\
S$_2$ & $-$1.39    & $-$2.99  & 0.06     & 0.16\\
\hline
\end{tabular}
\end{table}

\begin{figure*}[t!]
\centering
\includegraphics[trim=30 525 30 100,clip=true,width=\textwidth]{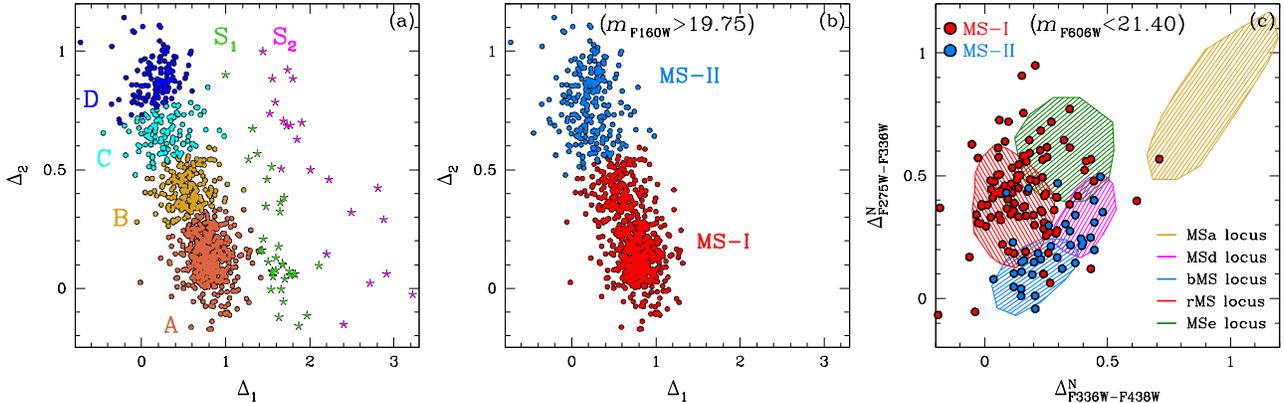}
\caption{(a) Reproduction of panel (c) of Fig.~\ref{fig:4}. (b)
  similar to panel (a), but for MS-I and MS-II stars below the MS knee
  ($m_{\rm F160W}<19.75$).  The brighter ($m_{\rm F606W}>21.4$) stars
  of MS-I and MS-II are also measured in F275W, F336W and F438W. We
  applied to these bright stars the same transformations used in
  \citet{2017ApJ...844..164B} to derive the $\Delta^{\rm N}_{\rm
    F275W-F336W}$ versus $\Delta^{\rm N}_{\rm F336W-F438W}$ chromosome
  map (panel c, see also their Fig.~10). On this, plane we highlighted
  the regions (colored shaded areas) occupied by the main population
  groups of \citet{2017ApJ...844..164B}:\ MSa in yellow, MSd in
  magenta, bMS in azure, rMS in red, MSe in green. The brighter MS-I
  and MS-II stars are shown as red and blue solid circles,
  respectively. We find MS-I stars mainly overlapped to the rMS and
  MSe loci, and MS-II stars to the bMS and MSd loci.}
\label{fig:7}
\end{figure*}

We have defined \mtan\ so that it is positive for a counter-clockwise
rotation in a righthanded Cartesian system in the plane of the
sky. Since \wcen\ is also rotating counter-clockwise, populations with
excess rotation have positive $\langle$\mtan$\rangle$, while the
opposite is true for populations rotating more slowly. This implies
that 1G stars, which are also characterized---on average---by a skewed
\mtan\ distribution, are rotating faster than 2G stars, which is the
main result of this Section. The difference is $0.084 \pm
0.017$\,\masyr (i.e., $\sim 2$\,\kms). For comparison, the overall
mean rotation velocity at the position of field F1 is believed to be
$\sim 5$\,\kms\ (e.g., \citealt{2006A&A...445..513V}). A direct
  measurement of the differential rotation of \wcen\ as a function of
  radius goes beyond the scope of the present work, but will be the
  subject of a stand-alone paper in this series once all the data of
  the remaining fields F0, F2, and F3 have been collected.

\subsection{A note on white dwarfs}\label{ss:3.5} 

Our PM catalog includes 29 relatively bright WD stars (see, e.g.,
panel (k) of Fig.~\ref{fig:3}).  Their velocity dispersion, computed
over the entire field are as follows:\ $\sigma_\mu=8.39\pm0.88$\,\kms,
$\sigma_{\rm rad}=8.98\pm1.32$\,\kms, and $\sigma_{\rm
  tan}=7.78\pm1.17$\,\kms. The resulting deviation from
radial-to-tangential isotropy is $-0.13\pm 0.18$. These values are
consistent with those obtained for the mPOPs on the MS of the
cluster. We also computed the median WD motion along the
$\alpha\cos\delta$, $\delta$, radial and tangential directions. All
these values are listed in Table~\ref{tab:2}. Given the large errors
due to small-number statistics, nothing definitive can be said about
the level of anisotropy of WDs in our field F1, or if WDs are rotating
faster or slower than MS stars. We will briefly return on the
kinematics of WDs at the end of Sect.~\ref{sec:4}.

\subsection{Who's who?}\label{ss:3.6} 

In this section, we want to see if it is possible to connect the six
populations identified below the MS knee (Paper~I) to the five
population groups (or even directly to the 15 distinct populations)
analyzed by \citet{2017ApJ...844..164B} on the upper MS in the core of
the cluster.  The meta-groups MS-I and MS-II play a pivotal role:\
since they extend through the entire MS at our disposal, they share
over $\gtrsim 2$ \mv\ magnitudes with the six populations of Paper~I
near the faint end of the MS, and can make use of F275W, F336W and
F438W filters in a $\sim 1$ \mv\ magnitude bin near the bright limit
of our PM catalog, so that we can apply the same selections criteria
used by \citet{2017ApJ...844..164B} to isolate the five population
groups in the cluster's core.

Panel (a) of Fig.~\ref{fig:7} is a replica of panel (c) of
Fig.~\ref{fig:4}, with the six populations of Paper~I.  On the same
plane in panel (b), we color-coded MS-I and MS-II below the MS knee in
red and azure, respectively. Clearly, MS-I stars are made up of
populations A and B, and MS-II stars are constituted by population C
and D stars. Neither the MS-I nor the MS-II overlaps with populations
S$_1$ and S$_2$, because of the way these stars were selected in
Paper~I.

We applied to MS-I and MS-II stars in the brightest $\sim 1$ \mv\
magnitude bin the same selections criteria of
\citet{2017ApJ...844..164B} to reproduce the $\Delta^N_{\rm
  F275W-F336W}$ versus $\Delta^N_{\rm F336W-F438W}$ chromosome map of
their Fig.~10. In panel (c) of Fig.~\ref{fig:7} we show this
chromosome map for the bright MS-I and MS-II stars in our PM catalog.

The shaded regions highlight the loci of the bulk of the five
populations groups identified in \citet{2017ApJ...844..164B}. On this
panel only, the color of the shaded regions refer to the colors
adopted by \citet{2017ApJ...844..164B} to distinguish the five
population groups:\ MSa in yellow, rMS in red, bMS in azure, MSd in
magenta, and MSe in green. MS-I and MS-II stars are shown as red and
blue solid circles, respectively. MS-I stars mostly overlap with the
bMS and MSd loci, while rMS stars mostly overlap with the rMS and MSe
loci.

From the analysis of \citet{2017ApJ...844..164B}, only $\sim 3.5$\% of
the bright MS stars in the core of the cluster are MSa stars. In
addition, studies of the radial distribution of RGB stars
\citet{2009A&A...507.1393B} show that the relative number of RGBa
stars (the progeny of the MSa) over that of the metal-poor RGB group
(the progeny of MS-I stars) halves from the cluster center out to
$\sim 17^\prime$ where our field F1 lies. There are a total of 137
stars plotted in panel (c) of Fig.~\ref{fig:7}, and only one MS-I star
is within the MSa locus. This star could actually be a MSa star, or it
could be an outlier. Either way, population abundances and radial
distributions tell us that we should not expect to have more than 1--2
MSa stars in panel (c), which is what we observe.

MS-I stars in panel (c) are mostly on the rMS locus, while in panel
(b) they are mostly located where population A stars are. This implies
that population A stars belong to the same population group as rMS
stars, while population B stars are more likely associated with the
MSe.

The identification of MS-II stars is less secure. Population D stars
are slightly more abundant than population C stars in panel (a), and
MS-I stars in panel (c) are slightly more abundant around the bMS
locus rather than around the MSd locus. It is tempting to associate
the population C to the MSd group, and the population D to the bMS
group.

We are left with populations S$_1$ and S$_2$. Statistically speaking,
either one can be associated to the MSa. Note, however, that the MSa
is the most extreme population group of the cluster in terms of
chemical abundance (${\rm [Fe/H]}\simeq -0.7$ (e.g.,
\citealt{1995ApJ...447..680N, 2002ApJ...568L.101P,
  2008ApJ...681.1505J, 2009ApJ...698.2048J, 2010IAUS..268..183M,
  2011ApJ...731...64M}), and possibly He abundance up to $Y=0.40$
(\citealt{2004ApJ...612L..25N, 2010AJ....140..631B,
  2017ApJ...844..164B}, Paper~I). Population S$_2$ is the most extreme
in terms of anisotropy (Fig.~\ref{fig:5} and Table~\ref{tab:2}) and
possibly differential rotation (Fig.~\ref{fig:6} and
Table~\ref{tab:3}), so it is tempting to associate the
kinematically-extreme S$_2$ stars to the chemically-extreme MSa.

Kinematically, the population S$_1$ looks more similar to MS-I stars
than to MS-II stars, so that S$_1$ stars should belong to either the
rMS or the MSe groups. \citet{2017ApJ...844..164B} argued that the
four subpopulations that form the MSe group could actually be split
into two, with the two least-populated MSe3 and MSe4 subpopulations
forming a stand-alone group. MSe3 and MSe4 stars account for just
about 3\% of MS stars in the core of the cluster, and S$_1$ stars
account for 2.3\% of the stars below the MS knee. We tentatively
associate S$_1$ stars to the MSe3 and MSe4 subpopulations. 

\section{State of energy equipartition}\label{sec:4}

\begin{figure*}[th!]
\centering
\includegraphics[trim=30 325 50 100,clip=true,width=\textwidth]{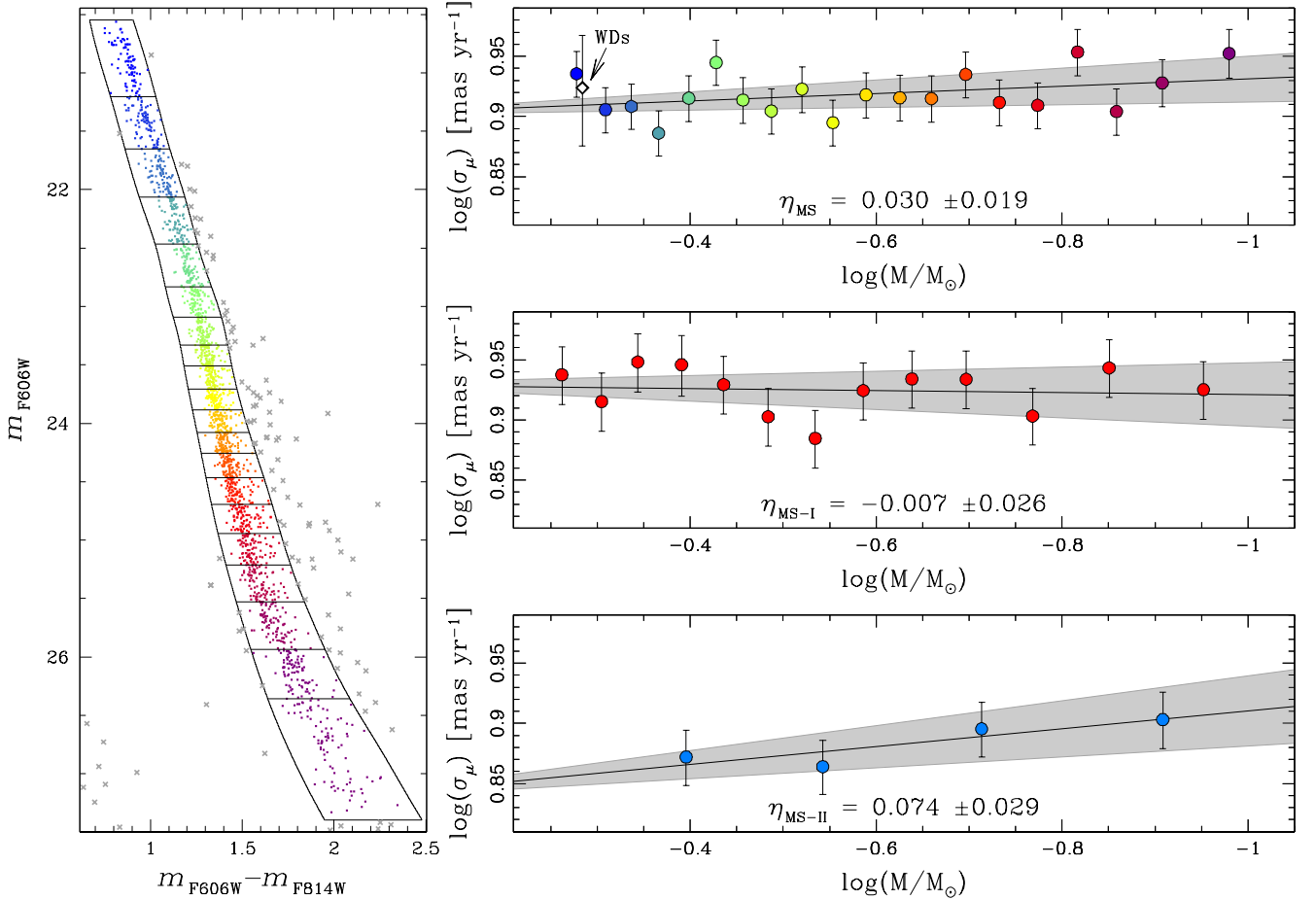}
\caption{On the left, the $m_{\rm F606W}$ versus $m_{\rm F606W}-m_{\rm
    F814W}$ CMD of the MS divided into 20 equally-populated bins,
  colored from blue to purple moving towards fainter stars. We used
  the same \citet{2008ApJS..178...89D} isochrones employed in Paper~I
  to transform stellar magnitudes into solar masses, weighting for the
  contribution of MS-I and MS-II stars in each bin. The top-right
  panel shows the combined velocity-dispersion profile of these bins
  as a function of stellar mass in a $\log$-$\log$ plane. For
  completeness, we also show the location of WDs stars (white diamond)
  with assumed average mass of 0.52\,$\mathrm{M}_\odot$. The slope of
  the straight-line fit to the MS points provides an estimate of the
  stage of energy equipartition of the cluster in our field. We find
  $\eta_{\rm MS}\simeq 0.03$. We also computed the state of energy
  equipartition separately for MS-I (red, middle-right panel) and
  MS-II (azure, bottom-right panel) stars. See the text for details.}
\label{fig:8}
\end{figure*}

By and large, GCs are assumed to evolve over many two-body relaxation
times towards a state of energy equipartition, for which the velocity
dispersion scales with stellar mass as $\sigma_\mu \propto m^{-\eta}$,
with $\eta=0.5$ (e.g., \citealt{1969ApJ...158L.139S,
  1987degc.book.....S}). Recently, \citet{2013MNRAS.435.3272T} and
\citet{2016MNRAS.458.3644B} have shown that this simple picture is
incorrect. In particular, \citet{2013MNRAS.435.3272T} used $N$-body
simulations to show that GCs reach a maximum value $\eta_{\rm max}
\approx 0.15$ in the core, with $\eta$ eventually asymptotically
declining to the value $\sim 0.08$. In the cluster outskirts, the
energy-equipartition indicator $\eta$ slowly evolves from zero to
$\sim 0.08$ (see, e.g., Figs.~6 and 7 of
\citealt{2013MNRAS.435.3272T}). The difference between the old and the
new pictures can be understood as a consequence of the Spitzer
instability for two-component systems, extended by
\citet{1978ApJ...223..986V} to a continuous mass spectrum.

Concerning \wcen, \citet{2013MNRAS.435.3272T} analyzed the PM catalog
of \citet{2010ApJ...710.1032A} of MS stars in the core of cluster, and
found $\eta \approx 0.2$ for stars in the mass range $0.5\leq
\mathrm{M}_\odot \leq 0.8$.  \citet{2013MmSAI..84..140B} extended the
mass range in the same field down to $\sim 0.3\,\mathrm{M}_\odot$, and
found $\eta = 0.16\pm 0.05$. These values imply that the core of the
cluster is somewhere in between being in no energy equipartition and
in full equipartition, and close to the peak value $\eta_{\rm max}$.

To measure the state of energy equipartition in field F1 we
proceeded as follows. We divided the MS of the cluster into 20
equally-populated magnitude bins in $m_{\rm F606W}$ (about 90 stars in
each bin, left panel of Fig.~\ref{fig:8}). We color-coded each bin
from blue to yellow to purple moving from the brighter to the fainter
bin.

To convert magnitudes into stellar masses, we fitted the same two
\citet{2008ApJS..178...89D} isochrones the MS-I and MS-II meta-groups
that we had applied in Paper~I. MS-I stars are well fitted by an
isochrone with ${\rm [Fe/H]}=-1.7$ and primordial Helium abundance
(${\rm Y}=0.246$). MS-II stars are modeled with a ${\rm[Fe/H]}=-1.40$,
${\rm Y=0.40}$ isochrone.  For each magnitude bin, we computed the
median mass of MS-I and MS-II stars within. We noted that 25\% of MS
stars are MS-II stars, while 75\% are MS-I stars (or a MS-I/MS-II
ratio of 0.34, in full agreement with the bMS/rMS ratio of
$0.34\pm0.05$ computed by \citet{2009A&A...507.1393B} for bright MS
stars in the same field). We assigned a weight of 0.75 and of 0.25 to
the computed median masses of MS-I and MS-II stars in each bin, and
averaged them together.

A direct way to estimate the energy-equipartition indicator $\eta$ is
to fit a least-squares straight line to the computed \sm\ of the 20 MS
bins versus mass in a $\log$-$\log$ plane (top-right panel in
Fig.~\ref{fig:8}). Points are color-coded according to the magnitude
bins on the CMD.  We obtain $\eta_{\rm MS}=0.030\pm0.019$. The gray
shaded region shows the $\pm 1 \sigma$ error in the slope. MS stars in
field F1 are consistent with being in nearly \textit{no energy
  equipartition}, in agreement with the predictions of
\citet{2013MNRAS.435.3272T}.

\citet{2013MNRAS.435.3272T} describe a canonical $N$-body model for
the interpretation of equipartition in \wcen\ and other GCs. The value
$\eta = 0.16 \pm 0.05$ observed near the center of \wcen\ is roughly
the maximum $\eta$ value attained in this model. At the time when this
maximum $\eta$ is attained near the center, the model value of $\eta$
at a projected radius equal to that of field F1 is only $0.012$. This
is lower because the relaxation times are much longer in the outskirts
of the cluster. The observed value $\eta_{\rm MS} = 0.030 \pm 0.019$
is consistent with this at the 1$\sigma$ level. This provides an
observational validation of the radial dependence of the partial
energy equipartition predicted by $N$-body models.

In addition, we also computed the state of energy equipartition
separately for MS-I and MS-II stars. To do this, we lowered the number
of magnitude bins in order to keep the same number of stars (about 90)
in each bin. Because of saturation in the long F606W exposures, our
mass ranges are $0.1 \lesssim \mathrm{M}_\odot \lesssim 0.6$ for the
MS-I meta-group, and $0.1 \lesssim \mathrm{M}_\odot \lesssim 0.5$ for
the MS-II meta-group. Results are shown in the middle-right (MS-I) and
bottom-right (MS-II) panels of Fig.~\ref{fig:8}. MS-I stars are
consistent with being in no energy equipartition, while there is some
marginal evidence that MS-II stars have $\eta_{\rm MS-II}\gtrsim 0$.

As a side note, the velocity dispersion of MS stars we find (about 8
\kms) is consistent with that based on spectra of RGB stars, and
reported by \citet{2006A&A...445..513V} at the same cluster distance
(about 17$^\prime$). Because of the relatively short evolutionary time
scales of the sub-giant-branch and the RGB, RGB stars have the same
``kinematic'' mass of MS turn-off stars (or about $0.8\,
\mathrm{M}_\odot$). The fact that stars about a factor of two more
massive than the typical MS stars studied here ($\sim
$0.35\,$\mathrm{M}_\odot$) have the same velocity dispersion is a
further proof that \wcen\ is not in equipartition in our field.

Finally, \wcen\ hosts two distinct WD cooling sequences
(\citealt{2013ApJ...769L..32B}), easily identifiable on UV-based
CMDs. The blue WD cooling sequence is populated by the evolved stars
of the He-normal (1G) component ($\sim 0.55\,\mathrm{M}_\odot$ CO-core
DA objects), while the red WD sequence hosts the end products of the
He-rich, 2G populations ($\sim 0.46\,\mathrm{M}_\odot$ objects, of
which $\sim 10$\% are CO-core and $\sim 90$\% are He-core WDs). A
detailed analysis of the WD cooling sequences of the cluster in our
four fields F0--F3 will be the subject of a future paper in this
series. For now, we can provide insights about the 29 WDs we have
identified in field F1 (Sect.~\ref{ss:3.5}). These WDs are aligned on
a single sequence in the \mv\ versus \mvi\ CMD (see, e.g.,
Fig.~\ref{fig:3}k). If WDs follow the same radial gradient as MS
stars, than we expect a 1G/2G ratio of 0.34, which translates into an
average per-star WD mass of $\sim 0.52\,\mathrm{M}_\odot$. Given a
$\sigma_\mu=8.39\pm0.88$\,\kms, WD stars are found to follow the same
trend of MS stars (white diamond in the top-right panel in
Fig.~\ref{fig:8}), well within their measured errors. Given the large
error bars, even a large over- or under-estimate of their mass would
still put them on the fitted line. This suggests that also WDs in the
field are consistent with not being in energy equipartition.

\section{Conclusions}\label{sec:5}

As part of the ``\textit{HST} large programme of $\omega$~Centauri''
(GO-14118\,+\,GO-14662, PI: Bedin, L.~R.), we have computed
high-precision PMs of MS stars of the GC \wcen\ down to near the HBL
in one of the four fields imaged by the program. This field, located
at about 3.5 half-light radii from the cluster center, is the first
for which all observations have now been completed. Well-measured
stars in the field have a typical PM error of $\sim
10\,\mu$as\,yr$^{-1}$.

We used the same population selections defined in Paper~I to study the
internal kinematics of the MS-I (1G) and MS-II (2G) meta-groups, as
well as of the six populations identified below the MS knee using a
combination of optical and IR filters. We find no significant trends
of the velocity dispersion as a function of distance from the cluster
center within our field. All populations have similar velocity
dispersions along the radial (i.e., towards the cluster's center)
component of the motion, \mrad. The velocity dispersions along the
tangential component of the motion, \mtan, are instead significantly
different between MS-I and MS-II stars, and between the six
populations below the MS knee. These kinematic differences result in
1G stars (MS-I, populations A, B, and S$_1$) being isotropic or nearly
isotropic, while 2G stars (MS-II, populations C, D and, S$_2$) are
radially anisotropic.  These results are consistent with what was
found at large radial distances in two other massive GCs:\ 47~Tuc
(\citealt{2013ApJ...771L..15R}) and NGC~2808
(\citealt{2015ApJ...810L..13B}), and can be interpreted as 2G stars
slowly diffusing towards the cluster's outer regions preferentially on
radial orbits.

The \mtan\ distribution of \wcen\ stars is found to be slightly skewed
towards negative \mtan\ values.  Recently, \citet{2017arXiv171010666H}
also measured some degree of skewness in the \mtan\ distribution of
47~Tuc stars, and pointed towards differential rotation being the
cause of the observed skewness. We found 1G stars to be slightly but
significantly skewed, while 2G stars have a more symmetric
\mtan\ distribution. In addition, the median values
$\langle$\mtan$\rangle$ of each population indicate that 1G
stars must have a higher velocity rotation in the plane of the sky
than 2G stars.

We identified MS-I and MS-II over the entire $m_{\rm F606W}$ magnitude
range in our PM catalog, so that these population meta-groups share
over two magnitudes in common with the six populations below the MS
knee. On the bright end of the MS, we used the photometric information
of the shallower F275W, F336W, and F438W photometry and applied to
them the same selection criteria used by \citet{2017ApJ...844..164B}
to identify 15 distinct populations (organized into 5 main groups) in
the core of the cluster. Using MS-I and MS-II stars as
a common benchmark, we were able to link the six populations of
Paper~I to the 5 groups of \citet{2017ApJ...844..164B}. We
find:\ population A $\Longleftrightarrow$ rMS, B
$\Longleftrightarrow$ MSe, C $\Longleftrightarrow$ MSd, D
$\Longleftrightarrow$ bMS. The connection of populations S$_1$ and
S$_2$ to their brighter counterparts is less obvious. Using chemical
abundance and kinematic arguments, we tentatively associated
population S$_1$ to the MSe3 and MSe4 populations, and S$_2$ to the
MSa.

We estimated the degree of energy equipartition of MS stars in the
mass range $0.1 \lesssim \mathrm{M}_\odot \lesssim 0.6$. We find
$\eta_{\rm MS} = 0.030 \pm 0.019$. This value is consistent with the
$N$-body simulations of \citet{2013MNRAS.435.3272T}, in which GCs
reach at most only partial equipartition. We also estimated the level
of energy equipartition separately for 1G (MS-I) and 2G (MS-II)
stars. We found the former to have $\eta_{\rm MS-I}=-0.007\pm0.026$
(no equipartition), and for the latter a value $\eta_{\rm
  MS-II}=0.074\pm0.029$, only marginally greater than zero.

We make our astro-photometric catalog publicly available to the
astronomical community through the ApJ website.  A description of the
catalog is given in Sect~\ref{A1}.

\acknowledgments Based on observations with the NASA/ESA
\textit{Hubble Space Telescope}, obtained at the Space Telescope
Science Institute, which is operated by AURA, Inc., under NASA
contract NAS 5-26555.  AB , AJB, DA, ML, and JMR acknowledge support
from STScI grant GO-14118.  APM acknowledges support by the European
Research Council through the ERC-StG 2016 project 716082 `GALFOR'.
This work has made use of data from the European Space Agency (ESA)
mission Gaia (\url{https://www.cosmos.esa.int/gaia}), processed by the
Gaia Data Processing and Analysis Consortium (DPAC,
\url{https://www.cosmos.esa.int/web/gaia/dpac/consortium}). Funding
for the DPAC has been provided by national institutions, in particular
the institutions participating in the Gaia Multilateral Agreement.
The Digitized Sky Surveys were produced at the Space Telescope Science
Institute under U.S. Government grant NAG W-2166. The images of these
surveys are based on photographic data obtained using the Oschin
Schmidt Telescope on Palomar Mountain and the UK Schmidt
Telescope. The plates were processed into the present compressed
digital form with the permission of these institutions.

\vspace{5mm}
\facilities{HST(ACS, WFC3)}

\software{Fortran, SM}

\appendix

\section{Description of the astro-photometric catalogs}\label{A1}

The catalog presented here is divided into an astrometric file
containing stellar positions and PMs, and one file per filter with the
photometric information.  These files contain the same number of
ordered entries (one line per star) so that, e.g., the PM data of line
100 in the astrometric file and the photometry of line 100 of any of
the photometric files refer to the same star.

Table~{\ref{tab:a1} shows the first 10 lines of the photometric file
  relative to the F814W filter.  The $o$ parameter tells us the
  initial (i.e., before neighbor subtraction) ratio between the light
  within the fitting radius due to nearby neighbors and the light of
  the star. The quantities n$_{\rm f}$ and n$_{\rm u}$ record in how
  many single exposures a star was found, and how many single
  measurements were used to compute photometric quantities,
  respectively.  The local sky value around each star and its RMS are
  given in units of e$^-$ at the reference exposure time. To revert
  VEGA mag values back into instrumental magnitudes $m$ (at the same
  reference exposure time), the user can simply subtract the adopted
  calibration zero points, which are given in Table~\ref{tab:a3}.
  Instrumental magnitudes can then be transformed into fluxes $F$ in
  unit of electrons:\ $F=10^{-0.4\times m}$ at the reference exposure
  time. This way, the user can apply selections based on sigmas over
  the sky background.  A flag value of 99.99 is used for the magnitude
  RMS when photometry is based on only one measurement. A flag value
  of zero is used for all the other quantities when a star is not
  measured in a particular filter. Undertermined local sky values are
  flagged with a value of 0.0, together with the associated sky RMS.

An extract of the astrometric file is given in Table~\ref{tab:a2}. The
X and Y positions refer to our right-handed Cartesian master frame,
which has a pixel scale of 40 mas\,pixel$^{-1}$. The
values $\chi^2_{\rm X}$ and $\chi^2_{\rm Y}$ are intended as
\textit{reduced} $\chi^2$ values. The temporal baseline used to compute
PMs is indicated as $\Delta$time. The ID entries in the last column of
the Table are the internal IDs of the reduction process.

\begin{table*}[h!]
\centering
\caption{First ten lines of the photometric catalog for filter F814W.}\label{tab:a1}
\small{
\begin{tabular}{rrrrrccrr}
\hline\hline
\multicolumn{1}{c}{VEGA mag} & \multicolumn{1}{c}{RMS mag} & \multicolumn{1}{c}{\texttt{QFIT}} & \multicolumn{1}{c}{$o$} & \multicolumn{1}{c}{\texttt{RADXS}} & \multicolumn{1}{c}{n$_{\rm f}$}& \multicolumn{1}{c}{n$_{\rm u}$}& \multicolumn{1}{c}{local sky (e$^-$)}& \multicolumn{1}{c}{RMS sky (e$^-$)}\\
\hline
  25.7403& 99.9900& 0.8760&  2.00481&$-$0.0945& 1& 1&     0.0&     0.0\\
  25.0999&  0.0308& 0.9530&  0.53509&$-$0.0169& 2& 2&     0.0&     0.0\\
  21.8048&  0.0003& 1.0000&  0.00000&$-$0.0026& 2& 2&    82.5&    51.4\\
  20.8638&  0.0099& 0.9950&  0.00000&$-$0.0052& 2& 2&   212.6&   115.0\\
  25.7705&  0.0335& 0.8170&  0.02167&   0.1728& 2& 2&    14.0&     9.1\\
  23.2805&  0.0070& 0.9970&  0.00000&   0.0046& 2& 2&    23.5&    14.1\\
  19.7016&  0.0011& 1.0000&  0.00000&$-$0.0008& 2& 2&   582.2&   318.3\\
  22.0618&  0.0015& 0.9990&  0.00000&$-$0.0052& 2& 2&    72.6&    38.5\\
  21.0557&  0.0431& 0.9820&  0.00000&   0.0156& 2& 2&   154.8&    89.0\\
  24.8920&  0.0153& 0.8930&  0.00000&   0.0910& 2& 2&     0.0&     0.0\\
 $[\dots]$  &$[\dots]$&$[\dots]$&$[\dots]$&$[\dots]$&$[\dots]$&$[\dots]$&$[\dots]$&$[\dots]$\\
\hline
\end{tabular}}
\end{table*}

\begin{table}[h!]
\centering
\caption{Adopted VEGA-mag photometric zero points.}\label{tab:a3}
\small{
\begin{tabular}{cc}
\hline\hline
\multicolumn{1}{c}{WFC3 Filter} & \multicolumn{1}{c}{VEGA zero point}\\
\hline
F606W&33.5960 \\
F814W&32.2910 \\
F110W&26.1749\\
F160W&24.7457\\
\hline
\end{tabular}}
\end{table}

\newpage
\begin{sidewaystable}
\caption{First ten lines of the astrometric catalog of field F1.}\label{tab:a2}
\tiny{
\centering
\begin{tabular}{rrrrrrrrrrrrrrrrrr}
\hline\hline
\multicolumn{1}{c}{R.A.} & \multicolumn{1}{c}{Decl.} & \multicolumn{1}{c}{X} & \multicolumn{1}{c}{Y} & \multicolumn{1}{c}{$\!\!\!\Delta \mu_\alpha^{\rm r} \cos\delta$} & \multicolumn{1}{c}{$\sigma_{\mu_\alpha^{\rm r} \cos\delta}$}& \multicolumn{1}{c}{$\!\!\!\Delta \mu_\delta^{\rm r}$} & \multicolumn{1}{c}{$\sigma_{\mu_\delta^{\rm r}}$}& \multicolumn{1}{c}{$\chi^2_{\rm X}$} & \multicolumn{1}{c}{$\chi^2_{\rm Y}$} & \multicolumn{1}{c}{n$_{\rm f}$}& \multicolumn{1}{c}{n$_{\rm u}$}& \multicolumn{1}{c}{$\!\!\!\Delta$time}& \multicolumn{1}{c}{$\!\!\!\Delta \mu^{\rm c}_\alpha \cos\delta$} & \multicolumn{1}{c}{$\sigma_{\mu^{\rm c}_\alpha \cos\delta}$} & \multicolumn{1}{c}{$\!\!\!\Delta \mu^{\rm c}_\delta$} & \multicolumn{1}{c}{$\sigma_{\mu^{\rm c}_\delta}$} & \multicolumn{1}{c}{$\!\!\!$ID}\\
\multicolumn{1}{c}{deg}&\multicolumn{1}{c}{deg}&\multicolumn{1}{c}{pixel}&\multicolumn{1}{c}{pixel}&\multicolumn{1}{c}{$\!\!\!\!$\masyr$\!\!\!\!$}&\multicolumn{1}{c}{$\!\!\!\!$\masyr$\!\!\!\!$}&\multicolumn{1}{c}{$\!\!\!\!$\masyr$\!\!\!\!$}&\multicolumn{1}{c}{$\!\!\!\!$\masyr$\!\!\!\!$}&   &   &   &   & \multicolumn{1}{c}{yr} &\multicolumn{1}{c}{$\!\!\!\!$\masyr$\!\!\!\!$}&\multicolumn{1}{c}{$\!\!\!\!$\masyr$\!\!\!\!$}&\multicolumn{1}{c}{$\!\!\!\!$\masyr$\!\!\!\!$}&\multicolumn{1}{c}{$\!\!\!\!$\masyr$\!\!\!\!$} \\
\multicolumn{1}{c}{(1)}&\multicolumn{1}{c}{(2)}&\multicolumn{1}{c}{(3)}&\multicolumn{1}{c}{(4)}&\multicolumn{1}{c}{(5)}&\multicolumn{1}{c}{(6)}&\multicolumn{1}{c}{(7)}&\multicolumn{1}{c}{(8)}&\multicolumn{1}{c}{(9)}&\multicolumn{1}{c}{(10)}&\multicolumn{1}{c}{(11)}&\multicolumn{1}{c}{(12)}&\multicolumn{1}{c}{(13)}&\multicolumn{1}{c}{(14)}&\multicolumn{1}{c}{(15)}&\multicolumn{1}{c}{(16)}&\multicolumn{1}{c}{(17)}&\multicolumn{1}{c}{(18)}\\
\hline
201.43076135593&$-$47.69362075406& 2384.4761& 1573.5081&   0.07308&  0.22500&$-$0.57616&  0.14524&   2.7706&   1.1542& 17& 14& 15.12775&   0.04832&  0.22749&$-$0.66180&  0.14989&   $\!\!\!$48\\
201.43075950173&$-$47.69362012573& 2384.5884& 1573.5647&$-$0.56564&  0.19632&$-$0.24208&  0.17492&   3.1617&   2.5091& 17& 15& 15.12775&$-$0.59040&  0.19916&$-$0.32772&  0.17881&   $\!\!\!$49\\
201.38848444873&$-$47.69361315403& 4945.5029& 1574.6746&   0.08380&  0.01828&   0.31376&  0.01812&   0.6405&   0.6287& 21& 21& 15.12775&   0.05456&  0.03819&   0.34896&  0.04023&   $\!\!\!$53\\
201.41179992503&$-$47.69395514603& 3533.1177& 1543.8008&$-$0.49612&  0.01724&   4.82204&  0.01676&   0.6072&   0.5738& 19& 19& 15.12725&$-$0.47240&  0.03873&   4.86792&  0.03304&   $\!\!\!$60\\
201.38992925597&$-$47.69366611367& 4857.9800& 1569.9146&   4.54220&  0.44612&   6.40428&  0.32588&  11.0648&   5.9048& 18& 18& 15.12775&   4.54856&  0.44737&   6.42320&  0.32772&   $\!\!\!$66\\
201.38889075924&$-$47.69363046474& 4920.8897& 1573.1185&$-$4.67476&  0.04188&   6.80408&  0.03684&   0.6947&   0.5374& 17& 15& 15.12764&$-$4.69240&  0.05391&   6.83928&  0.05124&   $\!\!\!$67\\
201.41577053661&$-$47.69392905370& 3292.5891& 1546.0909&$-$0.52392&  0.20692&   0.01040&  0.19460&   1.0987&   0.9716&  8&  8&  2.00041&$-$0.52296&  0.21015&   0.08496&  0.19766&   $\!\!\!$68\\
201.41597236104&$-$47.69371429518& 3280.3579& 1565.4168&   0.20084&  0.01504&$-$0.02724&  0.01548&   0.3271&   0.3463& 19& 17& 15.12775&   0.17764&  0.03818&   0.01004&  0.03636&   $\!\!\!$72\\
201.41529419354&$-$47.69363345110& 3321.4375& 1572.7040&   3.83644&  0.01900&   6.18012&  0.02324&   0.8338&   1.2490& 24& 23& 15.13011&   3.79600&  0.03939&   6.20736&  0.04021&   $\!\!\!$74\\
201.40595793882&$-$47.69364086614& 3887.0039& 1572.1504&   3.32892&  0.20576&   7.00512&  0.29964&   5.6617&  12.0045& 16& 15& 15.12775&   3.36304&  0.20901&   7.04240&  0.30114&   $\!\!\!$81\\
$[\dots]$&$[\dots]$&$[\dots]$&$[\dots]$&$[\dots]$&$[\dots]$&$[\dots]$&$[\dots]$&$[\dots]$&$[\dots]$&$[\dots]$&$[\dots]$&$[\dots]$&$[\dots]$&$[\dots]$&$[\dots]$&$[\dots]$&$\!\!\!$$[\dots]$\\
\hline \multicolumn{18}{l}{\textbf{Note.} The superscripts ``$^{\rm
    r}$'' and ``$^{\rm c}$'' of columns (5), (6), (7), (8), and (14),
  (15), (16), (17), respectively, refer to raw ($^{\rm r}$) and
  corrected ($^{\rm c}$) PMs, as described in Sects.~\ref{ss:2.5} and
  \ref{ss:2.6}.}\\

\end{tabular}}
\end{sidewaystable}


\begin{thebibliography}{}

\bibitem[Anderson(1997)]{anderson97} Anderson, J., Ph.D.\ thesis, Univ.\ of
  California, Berkeley, 1997 

\bibitem[Anderson \& King(2006)]{ak06} Anderson, J., \& King,
  I. R. 2006, ACS/ISR 2006-01 (Baltimore, MD: STScI), available online
  at \url{http://www.stsci.edu/hst/acs/documents/isrs} 

\bibitem[Anderson et al.(2006)]{2006A&A...454.1029A} Anderson, J.,
  Bedin, L.~R., Piotto, G., Yadav, R.~S., \& Bellini, A.\ 2006, \aap,
  454, 1029

\bibitem[Anderson et al.(2008)]{2008AJ....135.2055A} Anderson, J.,
  Sarajedini, A., Bedin, L.~R., et al.\ 2008, \aj, 135, 2055 

\bibitem[Anderson \& Bedin(2010)]{2010PASP..122.1035A} Anderson, J.,
  \& Bedin, L.~R.\ 2010, \pasp, 122, 1035 

\bibitem[Anderson \& van der Marel(2010)]{2010ApJ...710.1032A}
  Anderson, J., \& van der Marel, R.~P.\ 2010, \apj, 710, 1032 

\bibitem[Bastian et al. (2013)]{bastian} Bastian N. et al. 2013,
  \mnras, 436, 2398

\bibitem[Bedin et al.(2004)]{2004ApJ...605L.125B} Bedin, L.~R.,
  Piotto, G., Anderson, J., et al.\ 2004, \apjl, 605, L125 

\bibitem[Bedin et al.(2008)]{2008ApJ...678.1279B} Bedin, L.~R., King,
  I.~R., Anderson, J., et al.\ 2008, \apj, 678, 1279-1291 

\bibitem[Bellini \& Bedin(2009)]{bb09} Bellini, A., \& Bedin,
  L.~R.\ 2009, \pasp, 121, 1419 

\bibitem[Bellini et al.(2009a)]{2009A&A...507.1393B} Bellini, A.,
  Piotto, G., Bedin, L.~R., et al.\ 2009, \aap, 507, 1393 

\bibitem[Bellini et al.(2009b)]{2009A&A...493..959B} Bellini, A.,
  Piotto, G., Bedin, L.~R., et al.\ 2009, \aap, 493, 959 

\bibitem[Bellini et al.(2010)]{2010AJ....140..631B} Bellini, A.,
  Bedin, L.~R., Piotto, G., et al.\ 2010, \aj, 140, 631 

\bibitem[Bellini et al.(2011)]{b11} Bellini, A., Anderson, J., \&
  Bedin, L.~R.\ 2011, \pasp, 123, 622 

\bibitem[Bellini et al.(2013)]{2013ApJ...769L..32B} Bellini, A.,
  Anderson, J., Salaris, M., et al.\ 2013, \apjl, 769, L32 

\bibitem[Bellini et al.(2013)]{2013MmSAI..84..140B} Bellini, A., van
  der Marel, R.~P., \& Anderson, J.\ 2013, \memsai, 84, 140

\bibitem[Bellini et al.(2014)]{2014ApJ...797..115B} Bellini, A.,
  Anderson, J., van der Marel, R.~P., et al.\ 2014, \apj, 797, 115 

\bibitem[Bellini et al.(2015a)]{2015ApJ...810L..13B} Bellini, A.,
  Vesperini, E., Piotto, G., et al.\ 2015, \apjl, 810, L13 

\bibitem[Bellini et al.(2015b)]{2015ApJ...805..178B} Bellini, A.,
  Renzini, A., Anderson, J., et al.\ 2015, \apj, 805, 178

\bibitem[Bellini et al.(2017a)]{2017ApJ...842....6B} Bellini, A.,
  Anderson, J., Bedin, L.~R., et al.\ 2017, \apj, 842, 6 

\bibitem[Bellini et al.(2017b)]{2017ApJ...842....7B} Bellini, A.,
  Anderson, J., van der Marel, R.~P., et al.\ 2017, \apj, 842, 7 

\bibitem[Bellini et al.(2017c)]{2017ApJ...844..164B} Bellini, A.,
  Milone, A.~P., Anderson, J., et al.\ 2017, \apj, 844, 164 

\bibitem[Bellini et al.(2017d)]{2017ApJ...844..167B} Bellini, A.,
  Bianchini, P., Varri, A.~L., et al.\ 2017, \apj, 844, 167 

\bibitem[Bender et al.(1994)]{1994MNRAS.269..785B} Bender, R., Saglia,
  R.~P., \& Gerhard, O.~E.\ 1994, \mnras, 269, 785

\bibitem[Bianchini et al.(2016)]{2016MNRAS.458.3644B} Bianchini, P.,
  van de Ven, G., Norris, M.~A., Schinnerer, E., \& Varri,
  A.~L.\ 2016, \mnras, 458, 3644 

\bibitem[Bohlin(2016)]{2016AJ....152...60B} Bohlin, R.~C.\ 2016, \aj,
  152, 60 

\bibitem[Boyles et al.(2011)]{2011ApJ...742...51B} Boyles, J.,
  Lorimer, D.~R., Turk, P.~J., et al.\ 2011, \apj, 742, 51 

\bibitem[Bulmer(1979)]{bulmer} Bulmer, M.~G., 1979, ``Principles of
  Statistics''. Dover.

\bibitem[Cramer(1997)]{cramer} Cramer, D., 1997, ``Basic Statistics
  for Social Research''. Routledge.

\bibitem[D'Ercole et al.(2008)]{dercole2008} D'Ercole A., Vesperini
  E., D'Antona F., McMillan, S.L.W., Recchi, S., 2008, \mnras, 391,
  825 

\bibitem[D'Souza \& Rix(2013)]{2013MNRAS.429.1887D} D'Souza, R., \&
  Rix, H.-W.\ 2013, \mnras, 429, 1887 

\bibitem[Decressin et al.(2007)]{decressin} Decressin, T., Meynet,
  G., Charbonnel C. Prantzos, N.,Ekstrom,S. 2007, \aa, 464, 1029 

\bibitem[Di Nino et al.(2008)]{dinino08} Di Nino, D., Makidon, R.~B.,
  Latto, M., et al.\ 2008, ACS-ISR 2008-03 (Baltimore, MD:\ STScI),
  available online at
  \url{http://www.stsci.edu/hst/acs/documents/isrs/}

\bibitem[Dickens \& Woolley(1967)]{1967RGOB..128..255D} Dickens,
  R.~J., \& Woolley, R.~v.~d.~R.\ 1967, Royal Greenwich Observatory
  Bulletins, 128, 255

\bibitem[Dotter et al.(2008)]{2008ApJS..178...89D} Dotter, A.,
  Chaboyer, B., Jevremovi{\'c}, D., et al.\ 2008, \apjs, 178, 89-101 

\bibitem[Ferraro et al.(2002)]{2002ApJ...573L..95F} Ferraro, F.~R.,
  Bellazzini, M., \& Pancino, E.\ 2002, \apjl, 573, L95

\bibitem[Gaia Collaboration et al.(2016a)]{2016A&A...595A...2G} Gaia
  Collaboration, Brown, A.~G.~A., Vallenari, A., et al.\ 2016, \aap,
  595, A2

\bibitem[Gaia Collaboration et al.(2016b)]{2016A&A...595A...1G} Gaia
  Collaboration, Prusti, T., de Bruijne, J.~H.~J., et al.\ 2016, \aap,
  595, A1

\bibitem[Heyl et al.(2017)]{2017arXiv171010666H} Heyl, J., Caiazzo,
  I., Richer, H., et al.\ 2017, \apj, 850, 186

\bibitem[Harris(1996)]{h96} Harris, W.~E.\ 1996, AJ, 112, 1487, 2010
  edition

\bibitem[Johnson et al.(2008)]{2008ApJ...681.1505J} Johnson, C.~I.,
  Pilachowski, C.~A., Simmerer, J., \& Schwenk, D.\ 2008, \apj, 681,
  1505-1523

\bibitem[Johnson et al.(2009)]{2009ApJ...698.2048J} Johnson, C.~I.,
  Pilachowski, C.~A., Michael Rich, R., \& Fulbright, J.~P.\ 2009,
  \apj, 698, 2048

\bibitem[King et al.(2012)]{2012AJ....144....5K} King, I.~R., Bedin,
  L.~R., Cassisi, S., et al.\ 2012, \aj, 144, 5 

\bibitem[Lee et al.(1999)]{1999Natur.402...55L} Lee, Y.-W., Joo,
  J.-M., Sohn, Y.-J., et al.\ 1999, \nat, 402, 55

\bibitem[Libralato et al.(2018)]{libra.p3} Libralato, M., Bellini, A.,
  Bedin, L.~R., et al.\ 2018, arXiv:1801.01502 (Paper~III)

\bibitem[Marino et al.(2010)]{2010IAUS..268..183M} Marino, A.~F.,
  Piotto, G., Gratton, R., et al.\ 2010, Light Elements in the
  Universe, 268, 183

\bibitem[Marino et al.(2011)]{2011ApJ...731...64M} Marino, A.~F.,
  Milone, A.~P., Piotto, G., et al.\ 2011, \apj, 731, 64

\bibitem[Meylan \& Mayor(1986)]{1986A&A...166..122M} Meylan, G., \&
  Mayor, M.\ 1986, \aap, 166, 122

\bibitem[Milone et al.(2017a)]{2017MNRAS.464.3636M} Milone, A.~P.,
  Piotto, G., Renzini, A., et al.\ 2017, \mnras, 464, 3636 

\bibitem[Milone et al.(2017b)]{2017MNRAS.469..800M} Milone, A.~P.,
  Marino, A.~F., Bedin, L.~R., et al.\ 2017, \mnras, 469, 800
  (Paper~I) 

\bibitem[Norris \& Da Costa(1995)]{1995ApJ...447..680N} Norris, J.~E.,
  \& Da Costa, G.~S.\ 1995, \apj, 447, 680 

\bibitem[Norris(2004)]{2004ApJ...612L..25N} Norris, J.~E.\ 2004,
  \apjl, 612, L25

\bibitem[Pancino et al.(2000)]{pancino00} Pancino, E., Ferraro, F.~R.,
  Bellazzini, M., Piotto, G., \& Zoccali, M.\ 2000, \apjl, 534, L83

\bibitem[Pancino et al.(2002)]{2002ApJ...568L.101P} Pancino, E.,
  Pasquini, L., Hill, V., Ferraro, F.~R., \& Bellazzini, M.\ 2002,
  \apjl, 568, L101 

\bibitem[Pancino et al.(2017)]{2017MNRAS.467..412P} Pancino, E.,
  Bellazzini, M., Giuffrida, G., \& Marinoni, S.\ 2017, \mnras, 467,
  412

\bibitem[Piotto et al.(2015)]{2015AJ....149...91P} Piotto, G., Milone,
  A.~P., Bedin, L.~R., et al.\ 2015, \aj, 149, 91 

\bibitem[Platais et al.(2003)]{platais03} Platais, I., Wyse, R.~F.~G.,
  Hebb, L., Lee, Y.-W., \& Rey, S.-C.\ 2003, \apjl, 591, L127

\bibitem[Renzini et al.(2015)]{2015MNRAS.454.4197R} Renzini, A.,
  D'Antona, F., Cassisi, S., et al.\ 2015, \mnras, 454, 4197 

\bibitem[Richer et al.(2013)]{2013ApJ...771L..15R} Richer, H.~B.,
  Heyl, J., Anderson, J., et al.\ 2013, \apjl, 771, L15 

\bibitem[Sollima et al.(2007)]{2007ApJ...654..915S} Sollima, A.,
  Ferraro, F.~R., Bellazzini, M., et al.\ 2007, \apj, 654, 915

\bibitem[Spitzer(1969)]{1969ApJ...158L.139S} Spitzer, L., Jr.\ 1969,
  \apjl, 158, L139

\bibitem[Spitzer(1987)]{1987degc.book.....S} Spitzer, L.\ 1987,
  Princeton, NJ, Princeton University Press, 1987, 191 p.,

\bibitem[Trenti \& van der Marel(2013)]{2013MNRAS.435.3272T} Trenti,
  M., \& van der Marel, R.\ 2013, \mnras, 435, 3272 

\bibitem[van de Ven et al.(2006)]{2006A&A...445..513V} van de Ven, G.,
  van den Bosch, R.~C.~E., Verolme, E.~K., \& de Zeeuw, P.~T.\ 2006,
  \aap, 445, 513 

\bibitem[van der Marel \& Franx(1993)]{1993ApJ...407..525V} van der
  Marel, R.~P., \& Franx, M.\ 1993, \apj, 407, 525

\bibitem[van der Marel \& Anderson(2010)]{2010ApJ...710.1063V} van der
  Marel, R.~P., \& Anderson, J.\ 2010, \apj, 710, 1063 

\bibitem[van Leeuwen et al.(2000)]{2000A&A...360..472V} van Leeuwen,
  F., Le Poole, R.~S., Reijns, R.~A., Freeman, K.~C., \& de Zeeuw,
  P.~T.\ 2000, \aap, 360, 472

\bibitem[Vishniac(1978)]{1978ApJ...223..986V} Vishniac, E.~T.\ 1978,
  \apj, 223, 986

\bibitem[Watkins et al.(2013)]{2013MNRAS.436.2598W} Watkins, L.~L.,
  van de Ven, G., den Brok, M., \& van den Bosch, R.~C.~E.\ 2013,
  \mnras, 436, 2598

\bibitem[Woolley(1966)]{1966ROAn....2....1W} Woolley,
  R.~V.~D.~R.\ 1966, Royal Observatory Annals, 2,

\bibitem[Zhou et al.(2017)]{2017AJ....153..243Z} Zhou, Y., Apai, D.,
  Lew, B.~W.~P., \& Schneider, G.\ 2017, \aj, 153, 243

\end{thebibliography}
\end{document}